\documentclass[aps,prd,onecolumn,preprintnumbers,nofootinbib,amssymb]{revtex4-2}
\usepackage{graphicx,color}
\usepackage{amsmath}
\usepackage{amssymb}
\usepackage{amsfonts}
\usepackage{bm}
\usepackage{subfigure}
\usepackage[colorlinks=true, pdfstartview=FitV, linkcolor=blue, citecolor=red, urlcolor=magenta, breaklinks=true]{hyperref}

\def\be{\begin{equation}}
\def\ee{\end{equation}}
\def\bea{\begin{eqnarray}}
\def\eea{\end{eqnarray}}

\begin{document}

\title{Weak Coupling Regime in Dilatonic $f\left(R,T\right)$ Cosmology}

\author{F. A. Brito}
\email{fabrito@df.ufcg.edu.br}
\affiliation{	Departamento de Física, Universidade Federal da Paraíba, Caixa Postal 5008, 58051-970 João Pessoa, Paraíba, Brazil}
\affiliation{Departamento de Física, Universidade Federal de Campina Grande, Caixa Postal 10071, 58109-970 Campina Grande, Paraíba, Brazil}

\author{C. H. A. B. Borges}
\email{carlos.borges@ifpb.edu.br}
\affiliation{Instituto Federal de Educação Ciência e Tecnologia da Paraíba (IFPB),
Campus Campina Grande - Rua Tranquilino Coelho Lemos, 671, Jardim Dinamérica I.}
\affiliation{Departamento de Física, Universidade Federal da Paraíba, Caixa Postal 5008, 58051-970 João Pessoa,
Paraíba, Brazil}

\author{J. A. V. Campos}
\email{a.campos@uaf.ufcg.edu.br}
\affiliation{Departamento de Física, Universidade Federal de Campina Grande, Caixa Postal 10071, 58109-970 Campina Grande, Paraíba, Brazil}

\author{F. G. Costa}
\email{francisco.geraldo@ifpb.edu.br}
\affiliation{Instituto Federal de Educação Ciência e Tecnologia da Paraíba (IFPB),
Campus Campina Grande - Rua Tranquilino Coelho Lemos, 671, Jardim Dinamérica I.}

\begin{abstract}
We consider $f(R,T)$ modified theories of gravity in the context of string theory inspired dilaton gravity. We deal with a specific model that under certain conditions describes the late time Universe in accord with observational data in modern cosmology and addresses the $H_0$ tension. This is done by exploring the space of parameters made out of those coming from the modified gravity and dilatonic charge sectors. We employ numerical methods to obtain several important observable quantities.

\end{abstract}


\date{\today}

\maketitle

\section{Introduction}

Modern observations in cosmology, performed from Type Ia Supernova (SNIa)\cite{riess-pearlmutter}, Large Scale Structure (LSS)\cite{lss}, Wilkinson Microwave Anisotropy Probe (WMAP)\cite{wmap} data, Cosmic Microwave Background (CMB)\cite{cbm} and Baryonic Acoustic Oscillations (BAO)\cite{bao} indicate that the expansion of the Universe has entered an accelerated phase. Furthermore, the same observational data show that $95\%$ of the matter and energy content of the Universe (when described in terms of a fluid effectively entering Einstein's equations) is in the form of unknown species called Dark Matter (DM) and Dark Energy (DE). 
General Relativity has always been consistent with observational data and one of the most recent examples is the detection of gravitational waves through LIGO \cite{ligo}. General Relativity analyzes the acceleration of the Universe based on dark energy, according to observations of type Ia supernovae, which counterbalances gravitational attraction.
The Dark Energy \cite{{pad.1,frie,martin,cald,silve,bamba,li,sami}} component ($\rho_{x}$) is characterized by a negative effective pressure, $p_{x}<-\rho_{x}/3$. The simplest candidate for such a dark energy is a positive cosmological constant $\Lambda$, but such an identification raises some difficult questions, such as why $\Lambda$ is so small (in particle physics) \cite{{peeb,pad.2,sahni}}. This is called the fine-tuning problem for $\Lambda$. Another important problem is why $\Lambda=\rho_{x0}$ in present epoch (where $\rho_{x0}$ is the present value of the dark energy density of the Universe, in Planck units). This problem is known as cosmic coincidence \cite{velt}. A promising way to solve the above problems is to introduce a single scalar field, dubbed the quintessence \cite{shinji}, whose potential goes asymptotically to zero. The potential associated with this scalar field tends to zero as the field goes to infinity after an infinite (very long) time.
On the other hand, we can try to reconcile the observational data with the acceleration of the Universe through modifications of the theory of gravity. One way to do this is to start with a modification of the Einstein-Hilbert action formulation using an arbitrary function of the Ricci scalar, $f(R)$ \cite{buch}. In this context,  we assume that at large scales the Einstein gravity model breaks down. Some specific types of $f(R)$ models have been proposed in the literature (see \cite{revsotirioufaraoni} and related references for a recent review). These theories acquired a lot of interest following the work by Starobinsky on cosmic inflation \cite{starobin}. The late-time cosmic acceleration of the Universe, in this context, was first explained by a natural modification, adding terms to the action that are proportional to $R^{n}$ \cite{carroll}. Quintessence issues by taking into account generic models with actions containing $f(R)$ terms were addressed in Ref.~\cite{capozzi}.
Quantum effects may lead to a generalization of $f(R)$ theories of gravity \cite{quantum effect}. In Ref.~\cite{harko} it was developed by Harko \textit{et al} an unusual coupling between matter and geometry, where the gravitational sector is given by an arbitrary function of the Ricci scalar and the trace $T$ of the energy-momentum tensor. This type of theory is known as $f(R,T)$ gravity. Cosmological and astrophysical consequences of $f(R,T)$  models have been explored in the last few years. For instance, in Ref.~\cite{flrw}, cosmological solutions for a perfect fluid in a spatially flat Friedmann-Lemàıtre-Robertson-Walker (FLRW) metric is investigated. Other studies on cosmological applications of $f(R,T)$ gravity including, inflationary scenario, can be found in the Refs~\cite{Min,Moraes,Shabani,Debnath,Bhattacharjee.1,Bhattacharjee.2,Gamonal}.
An essential component of all superstring models and consequently of the cosmological scenarios based on the string effective action is the dilaton field. This field controls the effective strength of all gauge couplings in the context of “grand-unified” models of all fundamental interactions \cite{witten-dilaton.1}. This coupling strengths may drive the Universe towards a phase of strong coupling, what can possibly preceding the standard decelerated evolution. The dilaton can also be geometrically interpreted as the effective “radius” of the eleventh dimension \cite{witten-dilaton.2} in the $M$-theory context.
The dilaton may control the inflationary dynamics and play a role in the generation of the primordial spectra of quantum fluctuations amplified by inflation. String theory dilaton may provide a natural implementation of the coupled quintessence scenario, provided the cosmological running of the {dilaton does not stop after entering the weak coupling regime $e^{\phi} \ll1$, as $\phi\to-\infty$}. We shall consider a particular scenario in which the dilaton approaches to zero as $t\to \infty $. This is possible with an exponentially suppressed (non-perturbative) potential. As we shall see, because of the loop corrections, the fields inside the matter action are in general non-minimally and non-universally coupled to the dilaton. This will render dilatonic charge densities that are fundamental to form the new space of parameters of the model.

In Sec.~\ref{1} we will present the formalism of a theory of gravity modified by $f(R,T)$, addressing the variation of the modified action while in Sec.~\ref{stringy} we will review a scenario of string theory {at low energy $\phi\to -\infty$ with a dilatonic field subject to an effective potential $\tilde{V}$}. In Sec.~\ref{dilaton-fRT} we will analyze a dilatonic cosmological theory in a scenario of modified gravity of type $f(R,T)$ for a homogeneous and isotropic FLRW universe. The Sec.~\ref{results} is dedicated to detail the evolution of cosmological relevant quantities through numerical methods. Our discussions and conclusions are, respectively, present in Sec.~\ref{discussions} and Sec.~\ref{conclu}.


\section{Gravitational field equations of $f\left(R,T\right)$
gravity}\label{1}

Let us first take the action given in \cite{harko}:
\begin{equation}\label{1.1}
	S=\dfrac{1}{2\kappa}\int d^{4}x f(R,T)\sqrt{-g}+\int d^{4}x{\cal L}_{m}\sqrt{-g},	
\end{equation}
where $f(R,T)$ is an arbitrary function of the Ricci scalar curvature $R=g^{\mu\nu}R_{\mu\nu}$ and $T=g^{\mu\nu}T_{\mu\nu}$ is the trace of the energy-momentum tensor, ${\cal L}_{m}$ is the Lagrangian density of matter and $\kappa=8 \pi G$.

Admitting the definition of the energy-momentum tensor, we can express it in such a way that the Lagrangian density of matter depends only on $g_{\mu\nu}$, that is,
\begin{equation}\label{key}
	T_{\mu\nu}=g_{\mu\nu}{\cal L}_{m}-2\dfrac{\partial{\cal L}_{m}}{\partial g^{\mu\nu}}.	
\end{equation}
By varying the action given in equation \eqref{1.1} in relation to $ g^{\mu\nu} $, we have the field equations given by
\begin{eqnarray}
	f_{R}(R,T)R_{\mu\nu}&+&g_{\mu\nu}\Box f_{R}(R,T)-\nabla_{\mu}\nabla_{\nu}f_{R}(R,T)\nonumber\\
	&+&f_{T}(R,T)(T_{\mu\nu}+\Theta_{\mu\nu})-\dfrac{1}{2}f(R,T)g_{\mu\nu}-8\pi T_{\mu\nu}=0,	
\end{eqnarray}
so that $(T_{\mu\nu}+\Theta_{\mu\nu})$ corresponds to the variation of the trace with respect to the metric tensor, with $ \Theta_{\mu\nu}\equiv g^{\alpha\beta}\;\dfrac{\delta T_{\alpha\beta}}{\delta g^{\mu\nu}} $ set in \cite{harko}. We will denote $f_{R}(R,T)$ and $f_{T}(R,T)$ being the derivatives of $f(R,T)$ with respect to the Ricci scalar curvature and the trace of the energy-momentum tensor, respectively.

From the definition of $\Theta_{\mu\nu}$ and using equation (\ref{key}), we have
\begin{equation}\label{theta tensor}
	\Theta_{\mu\nu}=-2T_{\mu\nu}+g_{\mu\nu}{\cal L}_{m}-2g^{\alpha\beta}\dfrac{\partial^{2}{\cal L}_{m}}{\partial g^{\mu\nu}\partial g^{\alpha\beta}}.
\end{equation}
In other words, $ \Theta_{\mu\nu} $ will depend on the Lagrangian of matter, that is, this can refer to the case of the electromagnetic field, the massless scalar field, the case of the perfect fluid, among others.


\section{Stringy Cosmology}\label{stringy}

Let us now consider cosmological scenarios related to the effective action that comes from low energy string theory in which the dilaton field exerts influence in the dynamics of the Universe. We shall focus on the sector of the effective action, coming from a low energy string theory, given by the tensor field (the metric $\tilde{g}_{\mu\nu}$) and a scalar field (the dilaton $\phi$), where the tilde indicates that we are working on the \textit{string frame}, where the dilaton couples to the Ricci scalar and dilatonic dynamics in an explicit form --- See below.
Our starting point is the string-frame, low-energy, gravidilaton effective action, to lowest order in the $\alpha'$ expansion, but including dilaton-dependent loop and nonperturbative corrections, encoded in a few ‘‘form factors’’, due to the loop corrections $\psi(\phi)$ and $Z(\phi)$ \cite{Damour}. $V(\phi)$ is the effective dilaton potential. The model action is \cite{dil.mod.1}
\begin{equation}\label{sframe-1}
S=-\frac{M_{P}^{2}}{2}\int d^{4}x\sqrt{-\tilde{g}}\left[e^{-\psi(\phi)}\tilde{R}+\tilde{Z}(\phi)\left(\tilde{\nabla}\phi\right)^2+\frac{2}{M_{P}^{2}}\tilde{V}(\phi)\right]+ \tilde{S}_{\rm{m}}(\phi,\tilde{g},\rm{matter}). 
\end{equation}
We can discuss the phenomenology of the relic dilaton background by taking into account two possibilities. First, massive dilaton is gravitationally more \textit{strongly coupled} to macroscopic matter. In strong coupling limit $\phi\to\infty$ we assume that is possible to make an asymptotic Taylor expansion in inverse powers of the coupling constant $g_s^2=\exp(\phi)$ similar to what is done in the context of the “induced gravity”. { In these models the gravitational and gauge couplings saturate at small, but finite, values because of the very large number $N$ of fundamental gauge bosons presents in the loop corrections \cite{dil.mod.1}. By this assumption, we can write $\exp(-\psi(\phi))=c_1^2+b_1\exp(-\phi)$+${\cal O}(\exp(-2\phi))$, $Z(\phi)=-c_2^2+b_2\exp(-\phi)$+${\cal O}(\exp(-2\phi))$ and $\tilde{V}=V_0\exp(-\phi)$+${\cal O}(\exp(-2\phi))$, where $c_1^2$ and $c_2^2$ are dimensionless numbers. To be consistent with the tree-level relation $\lambda_P/\lambda_S = M_S/M_P=\exp(\phi/2)$ (for $d = 3$), in which
$M_P\simeq10 M_S$ as required by a consistent string unification in the context of gravitational and gauge interactions \cite{veneziano.lageN},
we have $c_1^2\sim c_2^2\sim 10^2$. On the other hand, very light (or massless) dilaton is weakly coupled to matter. In the rest of this paper we will focus our attention on this second  possibility, where we will consider that the dilaton is weakly coupled. In this regime we will admit that $\phi\to -\infty$ while $\exp(-\psi(\phi))=Z(\phi)=\exp(-\phi)$.} 

We can now characterize the dynamical evolution of the Universe with a metric minimally coupled to the dilaton. In this frame the string effective action is also minimally coupled to perfect fluid sources. Considering the lowest order $\alpha'$ we have \cite{dil.mod.1}
\begin{equation}\label{sframe}
S=-\frac{M_{P}^{2}}{2}\int d^{4}x\sqrt{-\tilde{g}}e^{-\phi}\left[\tilde{R}-\left(\tilde{\nabla}\phi\right)^2+\frac{2}{M_{P}^{2}}\tilde{V}(\phi)\right]+ \tilde{S}_m(\phi,\tilde{g},\rm{matter}). 
\end{equation}
We can use a more convenient coordinate system so called \textit{Einstein frame} in terms of the metric $g_{\mu\nu}$ that is defined by a conformal transformation $\tilde{g}_{\mu\nu}=e^\phi g_{\mu\nu}$. In this frame the action can be written as
\begin{equation}\label{eframe}
S=-\frac{M_{P}^{2}}{2}\int d^{4}x\sqrt{-g}\left[R-\frac{1}{2}\left(\nabla\phi\right)^2+\frac{2}{M_{P}^{2}}\hat{V}(\phi)\right]+ S_{\rm{m}}(\phi,e^\phi g_{\mu\nu},\rm{matter}),  
\end{equation}
where we have defined
\begin{equation}\label{new pot}
\hat V =  e^{\phi}\tilde{V}.
\end{equation}
Because of the loop corrections, the fields appearing in the action $S_{\rm{m}}$ are generally non-minimally and non-universally coupled to the dilaton \cite{Taylor}. {The gravitational and dilatonic “charge densities”, $T_{\mu\nu}$ and $\sigma$}, are defined as
\begin{equation}\label{charges}
\frac{\delta S_{\rm{m}}}{\delta g^{\mu\nu}}=\frac{1}{2}\sqrt{-g}T_{\mu\nu},\;\;\;\;\frac{\delta S_{\rm{m}}}{\delta \phi}=\frac{1}{2}\sqrt{-g}\sigma.
\end{equation} 
When $\sigma\neq 0$, the effective gravidilaton theory is very different from a typical scalar-tensor gravity model of the Jordan-Brans-Dicke type.


\section{Cosmology in dilatonic $f(R,T)$ Gravity}\label{dilaton-fRT}

Let us now assume that the action in equation (\ref{sframe}) depends not only on $R$, but on a function $f(R,T)$, so that $R$ is the Ricci scalar curvature and $T$ is the trace of the energy-momentum tensor of the dilatonic field. Since we are interested in investigating the dynamics of the dilaton field in the presence of other sources, we can rewrite this gravidilaton action as
\begin{equation}\label{model RT}
	S=-\dfrac{M^{2}_{p}}{2}\int d^{4}x\sqrt{-g}f(R,T)+\dfrac{M^{2}_{p}}{2}\int d^{4}x\sqrt{-g}\left[\dfrac{1}{2}g^{\mu\nu}\nabla_{\mu}\phi\nabla_{\nu}\phi-\frac{2}{M^2_p}\hat{V}(\phi)\right]+S_{\rm{m}}(\phi,e^\phi g_{\mu\nu},\rm{matter}). 
\end{equation}

Let us now assume a homogeneous and isotropic Universe described by the Friedmann-Lemaitre-Robertson-Walker (FLRW) metric whose line element is written as
\begin{equation}
	ds^{2}=dt^{2}-a^{2}(t)\left( dr^{2}+r^{2}d\theta^{2}+r^{2}{\sin}^{2}\theta\, d\phi^{2}\right). 	
\end{equation}
In what follows we will also consider $M_{P}^{2}\equiv{1}/{8\pi G}=2$, except otherwise indicated. For this model the dilatonic Lagrangian density ${\cal L}^{\phi}$ is given by 
\begin{equation}\label{dil.lag}
{\cal L}^{\phi}=\dfrac{1}{2}g^{\mu\nu}\nabla_{\mu}\phi\nabla_{\nu}\phi-\hat{V}(\phi),
\end{equation}
where $\hat{V}(\phi)=e^{\phi}\tilde{V}(\phi)$ is the effective dilaton potential as previously defined. The scalar dynamics of the model is governed by the energy-momentum tensor of the dilaton field and is given by	
\begin{equation}
	T^{\phi}_{\mu\nu}=\nabla_{\mu}\phi\nabla_{\nu}\phi-g_{\mu\nu}\left[ \dfrac{1}{2}g^{\alpha\beta}\nabla_{\alpha}\phi\nabla_{\beta}\phi-\hat{V}(\phi)\right],
\end{equation}
from which we can write
\begin{equation}\label{eq 00}
	T^{\phi}_{00}=\dfrac{1}{2}\dot{\phi}^{2}+\hat{V}(\phi),\\
\end{equation}
\begin{equation}\label{eq 11}
	T^{\phi}_{ii}=\dfrac{a^{2}}{2}\dot{\phi}^{2}-a^{2}\hat{V}(\phi).
\end{equation}
By tracing the energy-momentum tensor of the dilatonic field we are left with
\begin{eqnarray}\label{trace}
	T^{\phi}&=&-\dot{\phi}^{2}+4\hat{V}(\phi).
\end{eqnarray}
The dilaton equation of motion is given by
\begin{equation}\label{dil.eq}
\ddot{\phi}+3 H\dot{\phi}+\frac{d\hat{V}}{d\phi}+\frac{1}{2}\sigma= 0.
\end{equation}
Following (\ref{charges}), $\sigma$ is the charge associated to the coupling between dilatonic field and the fluid that makes up the Universe. In the usual way, $H= \dot a /a$ is the Hubble parameter and a dot denotes differentiation with respect to the Einstein cosmic time. For this model
\begin{equation}\label{dens pres scalar}
\rho_\phi= \frac{1}{2}\dot\phi^2 +\hat V(\phi), ~~~~~~~
p_\phi= \frac{1}{2}\dot \phi^2 -\hat V(\phi), 
\end{equation}
are the energy density and pressure of the dilaton field.

Varying equation (\ref{model RT}) with respect to $g_{\mu\nu}$, we will obtain
\begin{equation}\label{modifield eistein eq1}
f_{R}(R,T)R_{\mu\nu}-\frac{1}{2}f(R,T)g_{\mu\nu}+(g_{\mu\nu}\Box-\nabla_{\mu}\nabla_{\nu})f_{R}(R,T)=\frac{1}{2} T_{\mu\nu}-f_{T}(R,T)T_{\mu\nu}-f_{T}(R,T)\Theta_{\mu\nu},
\end{equation}
where $T_{\mu\nu}=T^{f}+T^{\phi}$ is total energy-momentum tensor of the perfect fluid ($f$) that fills the Universe (baryonic matter, radiation and dark matter) plus the dilaton field. Using equation (\ref{theta tensor}), we can write
\begin{eqnarray}\label{eq theta 1}
	\Theta_{\mu\nu}=-2\left(T_{\mu\nu}^{\phi}+T_{\mu\nu}^{f}\right) +g_{\mu\nu}\left( {\cal L}^{\phi}+{\cal L}^{f}\right),
\end{eqnarray}
with $ {\cal L}^{\phi}$ and $T_{\mu\nu}^{\phi}$  given by  (\ref{dil.lag}) and (\ref{dens pres scalar}), respectively. We assume that the energy-momentum tensor of the matter and energy is given by $T_{\mu\nu}=(\rho+p)u_{\mu}u_{\mu\nu}-pg_{\mu\nu}$, with conditions $u_{\mu}u^{\mu}=1$ and $u^{\mu}\nabla_{\nu}u_{\mu}=0$ satisfied by the four-velocity $u^{\mu}$. Therefore, the Lagrangian of the perfect fluid becomes ${\cal L}^{f}=-p$. In this way, we can write the $(0-0)$ and $(i-i)$ components of $\Theta_{\mu\nu}$ as
\begin{eqnarray}\label{eq theta 1b}
	\Theta_{00}=-\frac{1}{2}\dot{\phi}^2-3\hat{V}-2\rho-p,\;\;\;\Theta_{ii}=-\frac{3}{2}a^2\dot{\phi}^2+3a^2\hat{V}-a^2p.
\end{eqnarray}
In the following we will consider a simple example of $f(R, T)$ theories in order to study the late time evolution of the Universe with the presence of a single dilatonic field.

\subsection{Model $ f(R,T)=R+\alpha T^{\phi} $}

This model was first studied by Harko \textit{et al}. in Ref. \cite{harko}. These authors  reproduce, in $f(R,T)$ context, a relativistically covariant model of interacting dark energy based on a specific action \cite{Poplawski}. Another interesting aspect of this choice is that the gravitational coupling becomes an ``effective time dependent coupling", depending on the derivative of $f(T)$ with respect to the trace \cite{harko}. In order to reconcile our model with the the evidences which support General Relativity, we will assume a modified gravity model by $f(R,T)=R+\alpha T$. Using the expression (\ref{modifield eistein eq1}), we have
\begin{equation}\label{modifield eistein eq2}
R_{\mu\nu}-\dfrac{1}{2}g_{\mu\nu}R=\left( \dfrac{1}{2}-\alpha \right)T_{\mu\nu}-\alpha\Theta_{\mu\nu}+\dfrac{1}{2}g_{\mu\nu}\alpha T.	
\end{equation}
Thus, using the definitions (\ref{eq 00}), (\ref{eq 11}), (\ref{trace}) and (\ref{eq theta 1}) in (\ref{modifield eistein eq2}), one can obtain $(0,0)$ and $(1,1)$ components of field equations in a FLRW Universe as being
\begin{equation}\label{1mod.eq1}
	3H^{2}=\left( \dfrac{1}{2}-\alpha \right) \dfrac{1}{2}\dot{\phi}^{2}+\left( \dfrac{1}{2}+4\alpha \right)\hat{V}(\phi)+\frac{1}{2}\left( 1+3\alpha\right)\rho-\frac{1}{2} \alpha p,	
\end{equation}
\begin{equation}\label{1mod.eq2}
	2\dot{H}+3H^{2}=-\left( \dfrac{1}{2}+3\alpha\right)\dfrac{1}{2}\dot{\phi}^{2}+\left( \dfrac{1}{2}+4\alpha \right)\hat{V}(\phi)-\frac{1}{2}\left( 1+3\alpha\right)p+\frac{1}{2} \alpha \rho.	
\end{equation}
The combination of (\ref{1mod.eq1}), (\ref{1mod.eq2}) and (\ref{dil.eq}) leads to the coupled conservation equations for the matter (baryonic and dark), radiation and dilaton energy density, respectively:
\begin{equation}\label{cons.bare}
(1+3\alpha)\dot{\rho}_{b}+3H(1+2\alpha)\rho_{b}+\alpha\left(12H\dot{\phi}^2+10\frac{d\hat{V}}{d\phi}\dot{\phi}\right) =0,
\end{equation}
\begin{equation}\label{cons.dark}
(1+3\alpha)\dot{\rho}_{d}+\left[3(1+2\alpha)H-\frac{1}{2}(1-2\alpha)Q\dot{\phi}\right]\rho_{d} +\alpha\left(12H\dot{\phi}^2+10\frac{d\hat{V}}{d\phi}\dot{\phi} \right)=0,
\end{equation}
\begin{equation}\label{cons.radiation}
\left(1+\frac{8}{3}\alpha\right) \dot{\rho}_{r}+4(1+2\alpha)H\rho_{r}+\alpha\left(12H\dot{\phi}^2+10\frac{d\hat{V}}{d\phi}\dot{\phi}\right)=0,
\end{equation}
\begin{equation}\label{cons.dil}
\dot{\rho}_{\phi}+6H[\rho_{\phi}-(1+8\alpha)\hat{V}]-10\alpha \frac{d\hat{V}}{d\phi}\dot{\phi}+\frac{1}{2}(1-2\alpha)\sigma\dot{\phi}=0.
\end{equation}
In the set of equations shown above we have separate the radiation, baryonic and non-baryonic (dark)
matter components of the cosmological fluid by setting
\begin{equation}\label{total.dens}
\rho=\rho_{m}+\rho_r=\rho_b+\rho_d+\rho_r,\;\;\;p_b=p_d=0, \;\; p_r=\frac{1}{3}\rho_r.
\end{equation}
We assume that ordinary matter and radiation have nearly metric couplings, i.e. that $\sigma_b$ and $\sigma_r$ vanish as $\phi\to -\infty $. This agrees, for instance, with the precision tests of Newtonian gravity \cite{Fisch}. In the dark matter sector we shall assume a specific model to Lagrangian density. For ``cold dark matter", one has a dilatonic charge $\sigma_d$ which gives us the relationship \cite{dil.mod.1,Gasperini:1999ne}
\begin{equation}\label{q formula}
Q(\phi)=\frac{\sigma_d}{\rho_d}=Q_0 \frac{e^{Q_0\phi}}{c^2+e^{Q_0\phi}}.
\end{equation}
For large enough values of the constant $c^2$, $Q(\phi)$ approaches to finite (non zero) values in the limit of weak coupling regime, that is, when $\phi\to -\infty$.

Finally, we shall specify the form of the effective (Einstein frame) dilaton potential by choosing the string frame potential as $\tilde{V}(\phi)=V_0$. This allows us to write, quite generically, the simplest potential
\begin{equation}\label{dil.potential}
\hat{V}(\phi)=V_0e^{\phi},
\end{equation}
where $V_0$ is a constant. This potential is in agreement with the assumption of exponential suppression at weak coupling regime.

It is convenient to parameterize the temporal evolution of all variables in terms of the logarithm of the scale factor, $\chi=$\;$\rm{ln}$ $(a/a_{i})$. In this relationship, $a_i$ corresponds to the initial scale factor. So, the Einstein equation (\ref{1mod.eq1}) and the dilaton equation (\ref{dil.eq}) can be written, respectively, as
\begin{equation}\label{friedmann_chi}
H^2=\frac{(1+3\alpha)(\rho_b+\rho_d)+(1+\frac{8}{3}\alpha)\rho_r+(1+8\alpha)\hat{V}}{6-\left(\frac{1}{2}-\alpha\right)\left( \frac{d\phi}{d\chi}\right)^2},
\end{equation}
\begin{equation}\label{dil.eq_chi}
2H^2 \frac{d^2\phi}{d\chi^2}+\left[(1+4\alpha)\left(\frac{1}{2}\rho_b+\frac{1}{2}\rho_d+\frac{1}{3}\rho_r\right)+(1+8\alpha)\hat{V}-2\alpha H^2\left(\frac{d\phi}{d\chi}\right)^2\right] \frac{d\phi}{d\chi}+2\frac{d\hat{V}}{d\phi}+Q\rho_d=0.
\end{equation}
The matter and radiation evolution equations become
\begin{equation}\label{cons.bare_chi}
(1+3\alpha)\frac{d\rho_b}{d\chi}+3(1+2\alpha)\rho_{b}+\alpha\left[12H^2\left(\frac{d\phi}{d\chi} \right)^2+10\frac{d\hat{V}}{d\phi}\frac{d\phi}{d\chi}\right] =0,
\end{equation}
\begin{equation}\label{cons.dark_chi}
(1+3\alpha)\frac{d\rho_d}{d\chi}+3(1+2\alpha)\rho_{d}-\frac{1}{2}(1-2\alpha)Q\rho_{d}\frac{d\phi}{d\chi}+\alpha\left[12H^2\left(\frac{d\phi}{d\chi} \right)^2+10\frac{d\hat{V}}{d\phi}\frac{d\phi}{d\chi}\right]=0,
\end{equation}
\begin{equation}\label{cons.radiation_chi}
\left(1+\frac{8}{3}\alpha\right)\frac{d\rho_r}{d\chi}+4(1+2\alpha)\rho_{r}+\alpha\left[12H^2\left(\frac{d\phi}{d\chi} \right)^2+10\frac{d\hat{V}}{d\phi}\frac{d\phi}{d\chi}\right]=0.
\end{equation}
Finally,  the dilaton conservation equation can be written as
\begin{equation}\label{cons.dil_chi}
\frac{d\rho_{\phi}}{d\chi}+6[\rho_{\phi}-(1+8\alpha)\hat{V}]-10\alpha \frac{d\hat{V}}{d\phi}\frac{d\phi}{d\chi}+\frac{1}{2}(1-2\alpha)\sigma\frac{d\phi}{d\chi}=0,
\end{equation}
which is equivalent to the equation (\ref{dil.eq_chi}). Finally we defined the useful density parameter $\Omega=\rho/\rho_c$, where $\rho_c$ is the critical energy density. Thus, the Eq.~\eqref{friedmann_chi} can be rewritten in terms of different density parameters for each component of the Universe
\begin{equation}
1= \Omega_r+\Omega_d+\Omega_b+\Omega_\phi,
\end{equation}
where 
\begin{equation}\label{parameter-densities}
\Omega_r=\frac{\rho_r}{3H^2} ,\qquad \Omega_d=\frac{\rho_d}{3H^2},\qquad \Omega_b=\frac{\rho_b}{3H^2}, \qquad \Omega_\phi=\frac{\rho_\phi}{3H^2},
\end{equation}
are the density parameters of radiation, dark matter, baryonic matter and dilaton field, respectively.

\subsection{Numerics}

We can solve the set of equations (\ref{cons.bare_chi})-(\ref{cons.dil_chi}) by numerical methods with equation \eqref{friedmann_chi} as a constraint on the initial data set. Here, we also recover the $8\pi G$ factor, that can be given in terms of the Planck mass $M_{p}\equiv 1/\sqrt{G} = 1.22\times10^{19}$ GeV. We shall use a potential of the form \eqref{dil.potential}, by  assuming $V_{0} = 2.65\times10^{-123}M_{p}^{4} $, as well as the $Q(\phi)$ formula \eqref{q formula} with $c^{2} = 10$ for different values of the charge $Q_{0}$. 

As we shall see shortly,  $(\alpha, Q_0)$ is the space of parameters that we will explore to address acceptable cosmological scenarios.
To ensure that at the current time $\chi = 0$ the values of the energy densities are in agreement with the data from current cosmological measurements, we adjust the initial values of the energy densities of radiation, dark matter, baryonic matter and dark energy respectively, as
$\rho_{ri}(\chi_{i}) = 8.58\times 10^{-93}M_{p}^{4}$, $\rho_{di}(\chi_{i}) = 5.28\times 10^{-98}M_{p}^{4}$, $\rho_{bi}(\chi_{i}) = 9.30\times 10^{-99}M_{p}^{4}$ and $\rho_{\phi i}(\chi_{i}) = 1.67\times 10^{-105}M_{p}^{4}$, starting the integration at $\chi_{i} = -20$ --- this corresponds to big bang nucleosynthesis (BBN) redshift $z_{\rm BBN}\sim 10^{9}$. Furthermore, the initial value for the dilaton is $\phi_{i} = 7\times10^{-8}M_{p}$. 



\section{Results}\label{results}

\subsection{The density parameters $\Omega_r$, $\Omega_d$, $\Omega_b$ and $\Omega_\phi$}

\begin{figure}[h]
		\centering
		\includegraphics[scale=0.4]{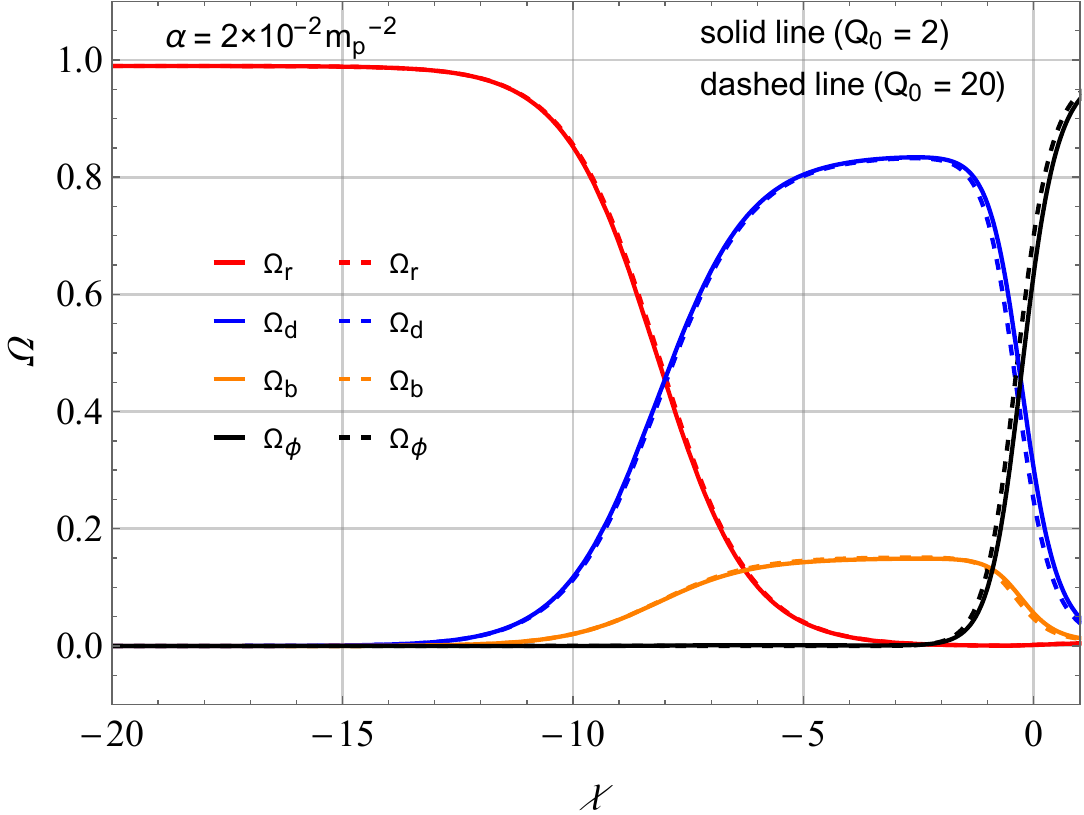}
		\caption{\small{The density $\Omega$ as a function of $\chi$ for fixed $\alpha=2\times10^{-2}m_{p}^{-2}$ and $Q_{0}=2$ and for $Q_{0}=20$.}}\label{Fig1}
\end{figure}

\begin{figure}[h]
	\includegraphics[scale=0.4]{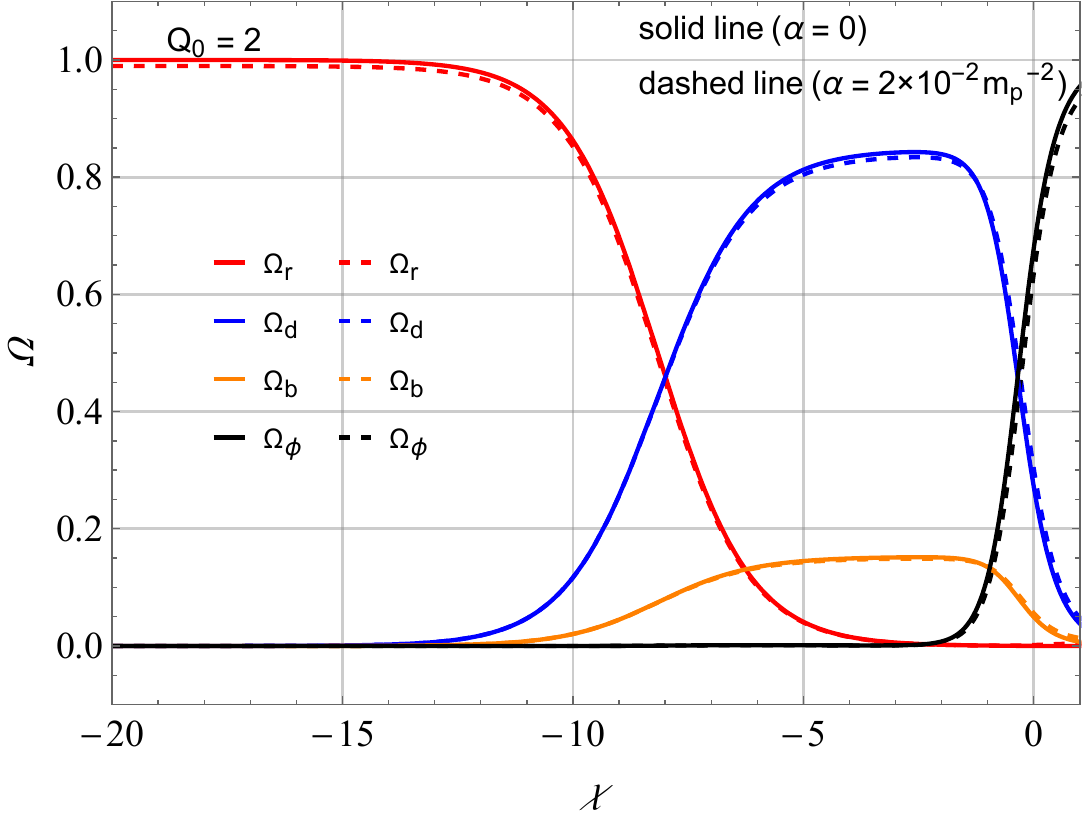}
	\includegraphics[scale=0.4]{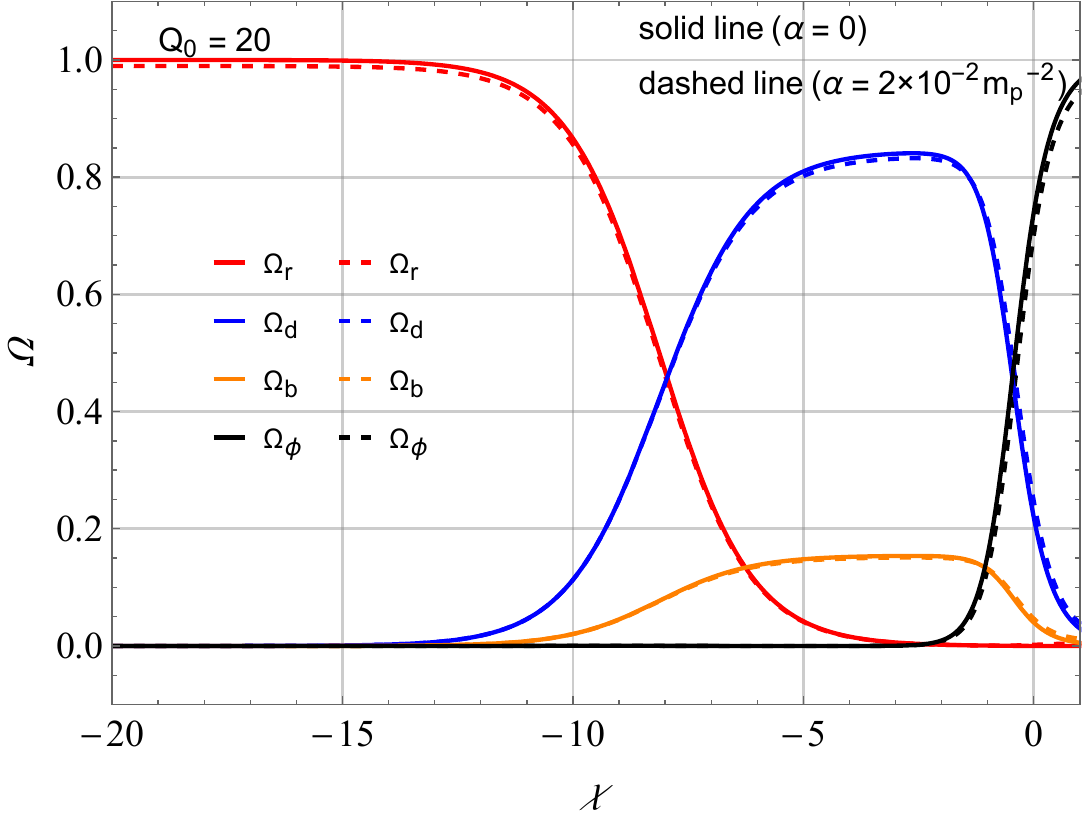}
	\caption{\small{({\it left panel}) The density $ \Omega $ as a function of $ \chi $ for fixed $ Q_{0}=2 $ and ({\it right panel}) $ Q_{0}=20 $ for $\alpha=0$ and $\alpha=2\times10^{-2}m_{p}^{-2}$.}}\label{Fig2}
\end{figure}

The density parameters \eqref{parameter-densities} are depicted in the Fig.~\ref{Fig1} and Fig.~\ref{Fig2}. The graph in Fig.~\ref{Fig1} expresses the behavior of the density parameter $\Omega$ of each
component as a function of $\chi$, by fixing $\alpha = 2 \times 10^{-2}m_p^{-2}$ and adjusting scenarios with
$Q_0 = 2$ and $Q_0 = 20$. Considering the radiation component $\Omega_r$, we note that it presents a degenerate evolution that starts constantly between the intervals $\chi\approx [-20, -13]$, decays quickly between $\chi\approx[-15, -4]$ and becomes null outside both intervals. In other words, $Q_0$ does not interfere with the evolution of $\Omega_r$.
Regarding the dark matter component $\Omega_d$, we have a degenerate behavior in the interval $\chi\approx[-20, -2]$ and a minimal influence of $Q_0$ on $\chi\approx[-2, 1]$, in order to present a slightly more accentuated decay with the decrease of $Q_0$. In this curve, the evolution begins with $\Omega_d=0$ in the intervals $\chi\approx[-20, -13]$ and a sharp increase in $\chi\approx[-13, -2]$ that precede the decay, there is also a degenerate maximum point with $\Omega_d(\chi=-2)\approx 0.9$.
For the baryonic matter component $\Omega_b$, we have an evolution with a degeneracy in the interval $\chi\approx[-20, -1]$ and a sufficiently small influence of $Q_0$ outside this interval. The behavior of $\Omega_b$ starts with a null value at $\chi\approx[-20, -13]$ followed by a gentle increase, remaining constant at its maximum point with $\Omega_b=0.15$  at $\chi\approx[-5, -2]$ and ends with a decay outside these ranges.
Regarding the dilaton component $\Omega_\phi$, we have a beginning with $\Omega_\phi=0$ that is independent of $Q_0$ in the interval $\chi\approx[-20, -2]$ and a significant elevation that grows smoothly with the increase in $Q_0$.
Relating the components $\Omega_r$, $\Omega_d$ and $\Omega_b$, we note that at $\chi\approx-13$, the first decreases while the second and third increase. On the other hand, at $\chi\approx-2$, $\Omega_r$ remains null and $\Omega_d$ and $\Omega_b$ decrease while $\Omega_\phi$ grows significantly, thus presenting an intersection with $\Omega_d$ at 
$\chi\approx-0.28$ and $\chi\approx-0.41$ for $Q_0 = 2$ and $Q_0 = 20$, respectively. We also have another point of intersection relating the components $\Omega_d$ and $\Omega_r$ given in $\chi\approx-8.01$ and $\chi\approx-7.97$ for $Q_0 = 2$ and $Q_0 = 20$, respectively. 

The graphs in Fig.~\ref{Fig2} ({\it left panel}) and Fig.~\ref{Fig2} ({\it right panel}) are similar to each other, however, in Fig.~\ref{Fig2} ({\it right panel}) we set $Q_0 = 20$ and adjust $\alpha$ for two values as in Fig.~\ref{Fig2} ({\it left panel}). Comparing these graphs with Fig.~\ref{Fig1}, we can observe that the increase in $Q_0$ has minimal influence on the radiation $\Omega_r$, dark matter $\Omega_d$ and baryonic $\Omega_b$ components. However, for the dilaton component $\Omega_\phi$, we note that there is a more pronounced rise in the respective evolution curve.

\subsection{The Hubble parameter $H(z)$}

\begin{figure}[h]
		\centering
		\includegraphics[scale=0.4]{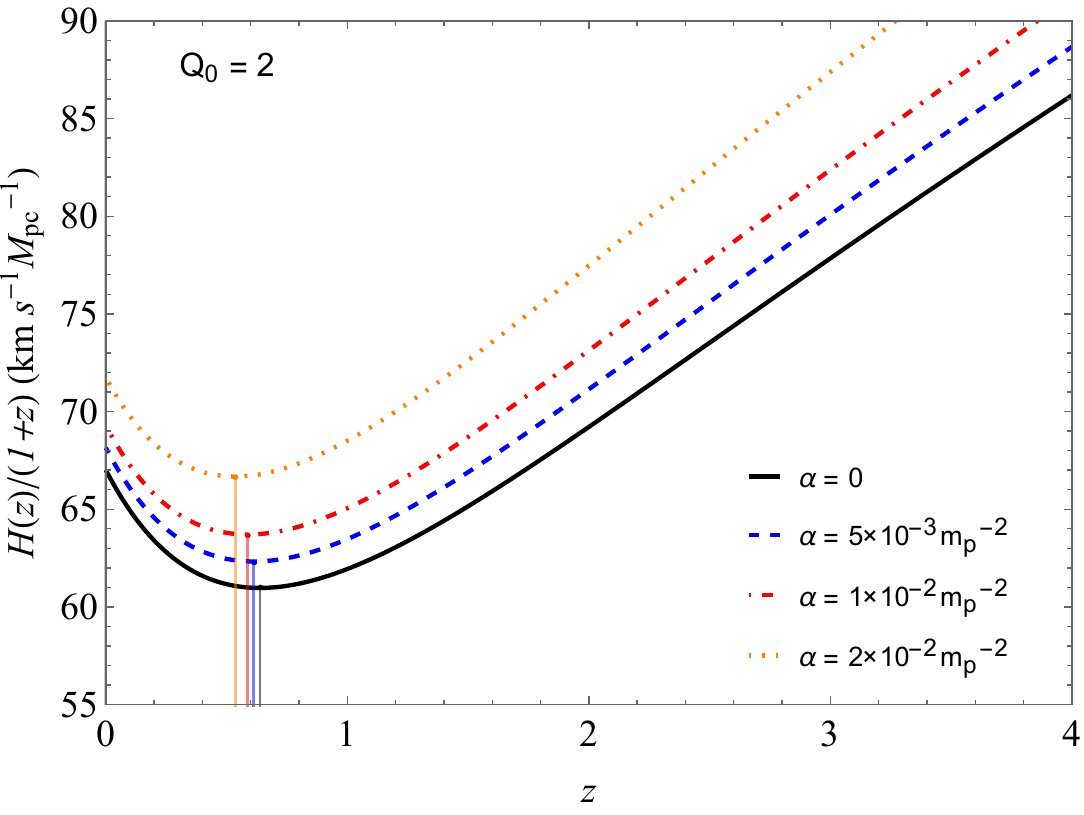}
		\caption{\small{The Hubble parameter $H(z)$ as a function of the redshift $z$ for fixed $Q_{0}=2$ and several values of $\alpha$.}}\label{Fig1-1}
\end{figure}

The graph in Fig.~\ref{Fig1-1} expresses the behavior of the Hubble parameter as a
function of redshift $z$, where we fixed $Q_0 = 2$ and explored different values for $\alpha$. In this
case, we can observe that in all values admitted for $\alpha$ we have a decay
followed by an increase. Note that the minimum point of each curve has a redshift $z$ that decreases with
the increase in $\alpha$, that is, for $\alpha=0$, $\alpha=5\times10^{-3}m_p^{-2}$, $\alpha = 1\times10^{-2}m_p^{-2}$ and $\alpha=2\times10^{-2}m_p^{-2}$
we have $z\approx0.64$, $z\approx0.61$, $z\approx0.59$ and $z\approx0.54$, respectively.
On the other hand, the values of the Hubble parameter at $z=0$, i.e., $H_0$ increases with the growth of $\alpha$, assuming values between $H_0\approx67$ km s$^{-1}$ Mpc$^{-1}$ and $H_0\approx72$ km s$^{-1}$ Mpc$^{-1}$. 

\begin{figure}[h]
	\includegraphics[scale=0.4]{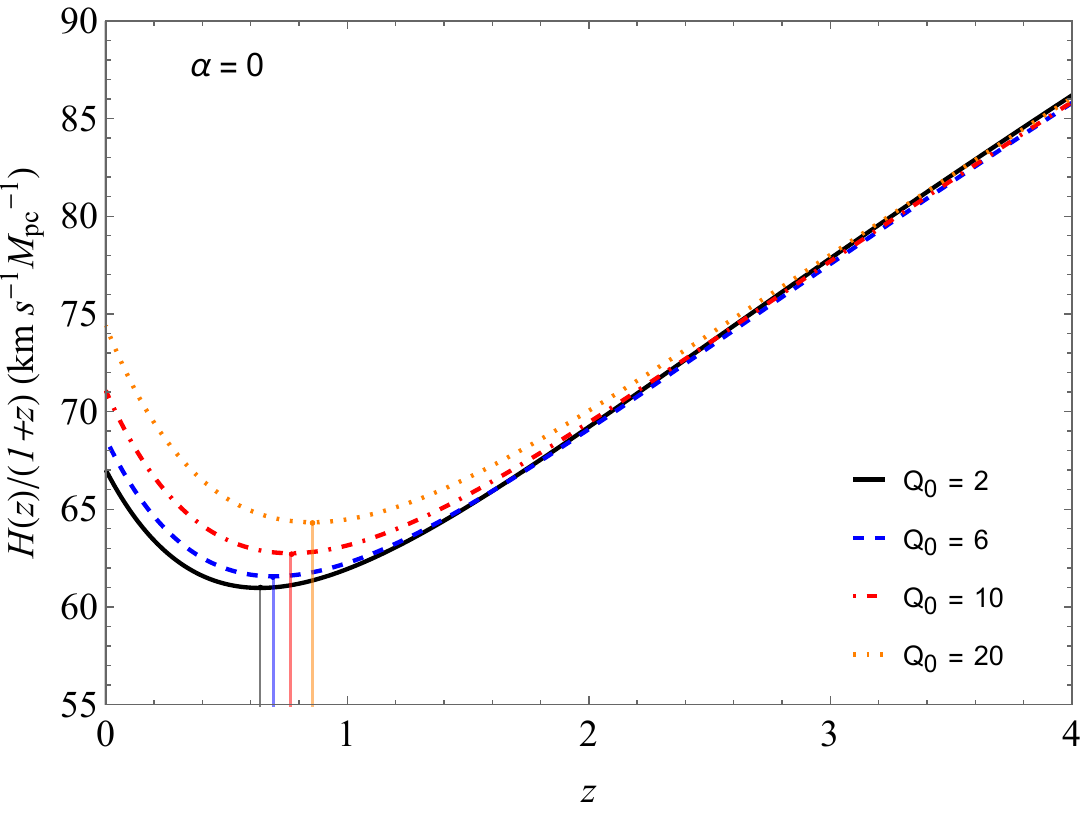}
	\includegraphics[scale=0.4]{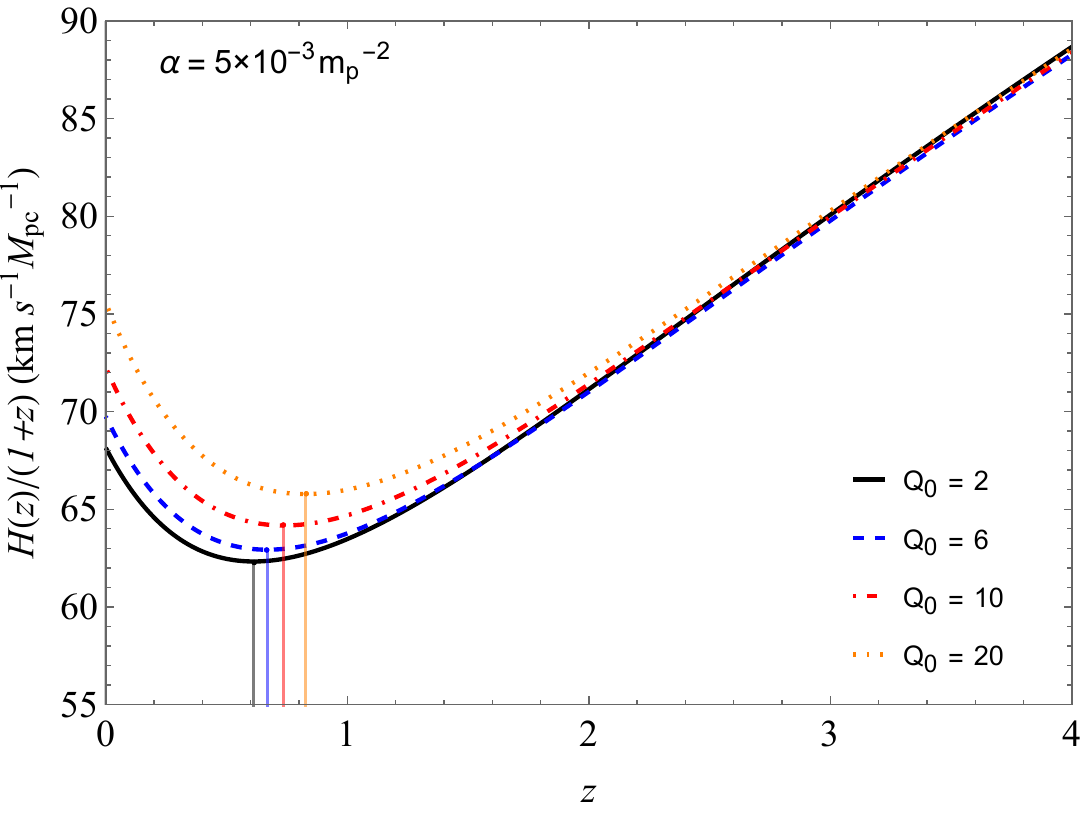}
	\caption{\small{({\it left panel}) The Hubble parameter $H(z)$ as a function of the redshift $z$ for fixed $\alpha=0$ and ({\it right panel}) $\alpha=5\times 10^{-3}m_p^{-2}$ and several values of $Q_0$.}}\label{Fig2-2}
\end{figure}

The graphs in Fig.~\ref{Fig2-2} show opposite behavior of the graph shown in Fig.~\ref{Fig1-1} with respect to the parameters, where they have interchanged their roles. More specifically, the Fig.~\ref{Fig2-2} ({\it left panel}) shows the behavior of the Hubble parameter as a function of redshift $z$ for different values of $Q_{0}$, for $\alpha=0$ fixed. Notice  that in all curves, we have a decay of $H$ followed by an increase in which there is a degeneracy starting at $z\approx2.6$. For $Q_{0}=2$, $Q_{0}=6$, $Q_{0}=10$ and $Q_{0}=20$ we have a minimum point at $z\approx 0 .64$, $z\approx 0.69$, $z\approx 0.76$ and $z\approx 0.85$, respectively. We also note that the values of $z$, referring to each point of minimum, increase with the increase of $Q_{0}$. It is also worth noting that the decay of $H$ to the minimum point becomes more pronounced with the increase of $Q_{0}$, as well as for smaller values of $Q_{0}$ we have slower decays. In this case,  the values of the Hubble parameter at $z=0$, i.e., $H_0$ assume values between $H_0\approx67$ km s$^{-1}$ Mpc$^{-1}$ and $H_0\approx74$ km s$^{-1}$ Mpc$^{-1}$. Finally, with respect to the graph in Fig.~\ref{Fig2-2} ({\it right panel}), we find that it is similar to the Fig.~\ref{Fig2-2} ({\it left panel}), but with $ \alpha=5\times10^{-3}m_{p}^{-2} $ fixed. In this case, we again have a decay followed by an increase that becomes degenerate at $z\approx2.4$. The curves assume minimum points whose redshift values are $z\approx0.61$, $ z\approx0.67$, $z\approx0.74$ and $z\approx0.82$ corresponding to $Q_{0}=2$, $Q_{0}=6$, $Q_{0}=10$ and $Q_{0}=20$ respectively. We can see that $H$ increases with the increase in $ \alpha$, and consequently, its minimum points take on greater values for $H$, however, the corresponding redshift values $z$ decrease. Thus, in this case,  the values of the Hubble parameter at $z=0$, i.e., $H_0$ assume values between $H_0\approx68$ km s$^{-1}$ Mpc$^{-1}$ and $H_0\approx75.5$ km s$^{-1}$ Mpc$^{-1}$. 

\subsection{The dilaton field}

\begin{figure}[h]
		\centering
		\includegraphics[scale=0.4]{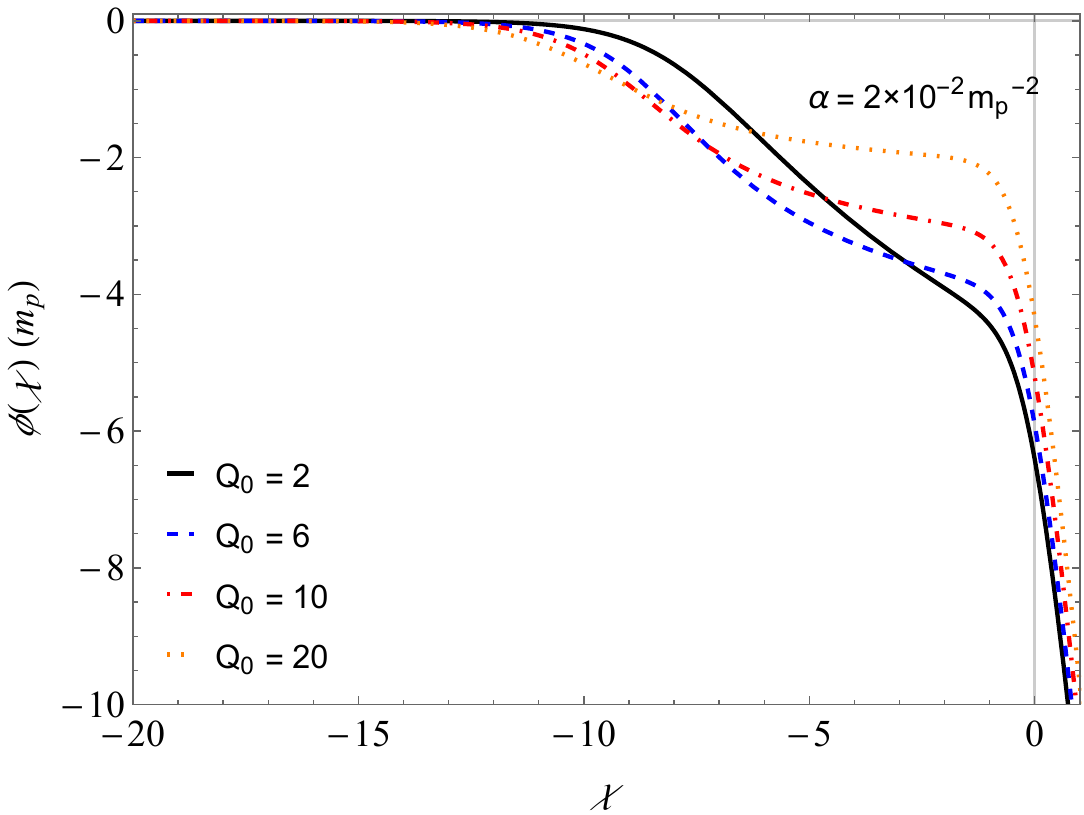}
		\caption{\small{The dilaton field $\phi$ as a function of $\chi$ for fixed $\alpha=2\times10^{-2}m_{p}^{-2}$ and several values of $Q_0$.}}\label{Fig3-1}
\end{figure}

The graph in Fig.~\ref{Fig3-1} details the behavior of $\phi$ as a function of $\chi$ assuming $\alpha=2\times10^{-2}m_{p}^{-2}$ being fixed and exploring different values for $Q_{0}$. In this case, there is a convergence of $\phi(\chi)=0$ for each $Q_{0}$ in the interval $\chi\approx[-20, -13] $, followed by a decay that ends in $\phi(\chi\approx1)=-10$. Notice that, in the range $ \chi\approx[-10, -1] $, $\phi(\chi)$ presents a decay that becomes more pronounced with the decrease of $Q_{0}$, that is, the growth of $Q_{0}$ reflects a slower decay to $\phi(\chi)$. This can be observed more clearly by analyzing the decay in which $Q_{0}=20$, where $\phi(\chi\approx-7.5)\approx-1.5$ and $\phi(\chi\approx-2)\approx-2 $, while for $Q_{0}=6$ we have $\phi(\chi\approx-8)\approx-1.5$ and $\phi(\chi\approx-7.3)\approx-2$. We also observe that from $\chi\approx-1$, we have an expressive decay that is independent of $ Q_{0}$.

\begin{figure}[h]
	\includegraphics[scale=0.4]{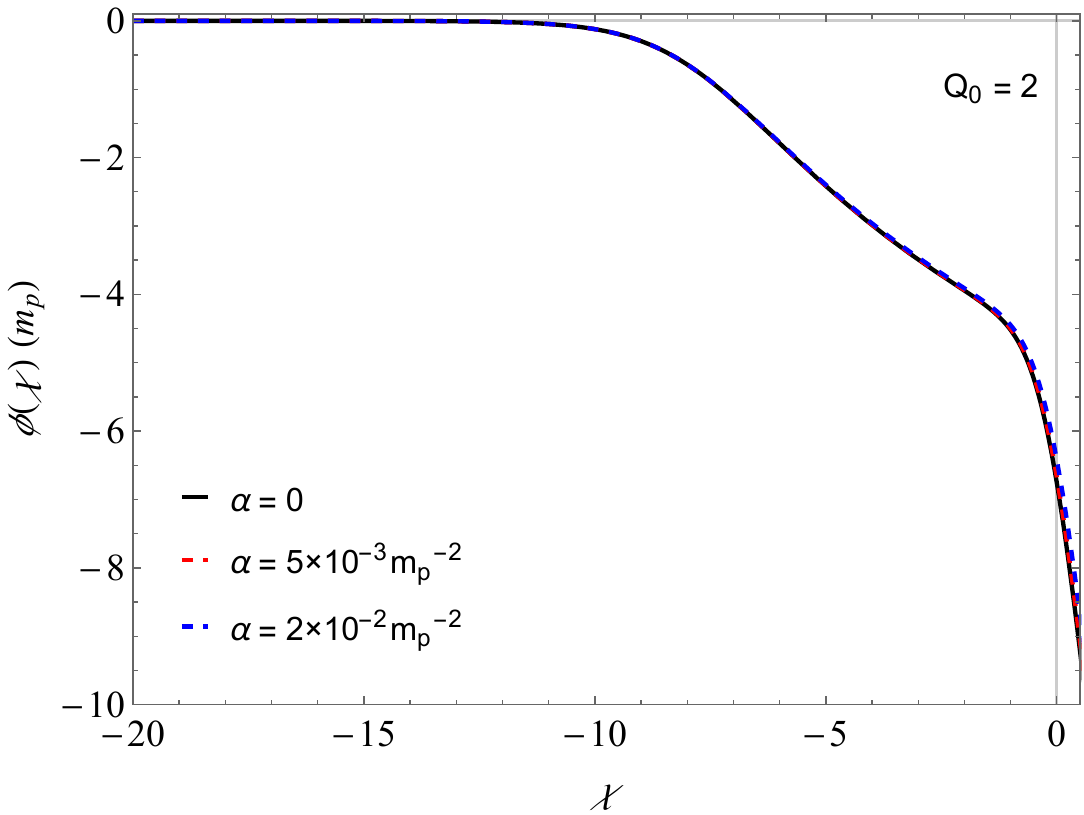}
	\includegraphics[scale=0.4]{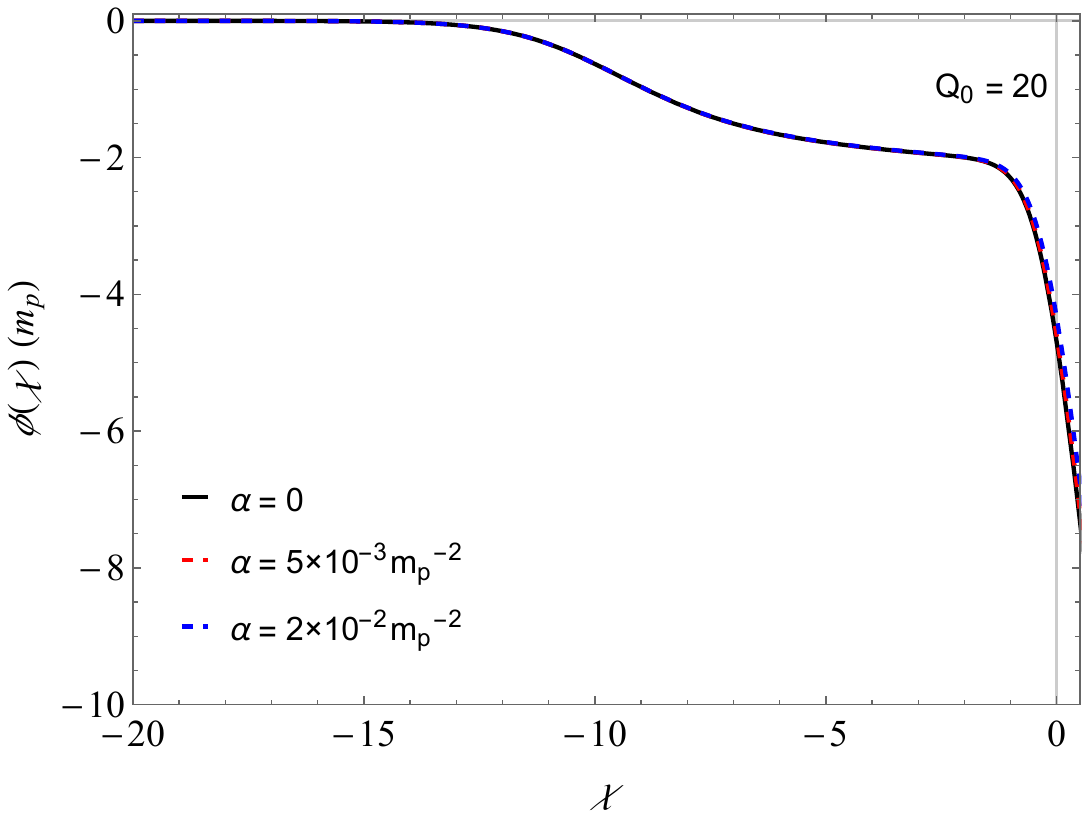}
	\caption{\small{({\it left panel}) The dilaton field $\phi$ as a function of $\chi$ for fixed $Q_0=2$ and ({\it right panel}) $Q_0=20$ and several values of $\alpha$.}}\label{Fig3-2}
\end{figure}

The graphs in Fig.~\ref{Fig3-2} are quite similar. In the first case ({\it left panel}) we fix $Q_{0}=2$ and explore three values for $\alpha$. As we can see, the variation of $\alpha$ presents a sufficiently small change in the behavior of $\phi(\chi)$ that decays in a degenerate way in the interval $\chi\approx[-20, -1]$, while for the range $\chi\approx[-1, 1]$ we have a decay that becomes smoothly slower with increasing $\alpha$. In the second case ({\it right panel}) we fix $Q_{0}=20$ and explore three values for $\alpha$. Furthermore,  we note that the increase in $Q_{0}$ contributes to a slower decay in the interval $\chi\approx[-20, -1]$.

\subsection{The running of $Q(\phi)$}

\begin{figure}[h]
		\centering
		\includegraphics[scale=0.4]{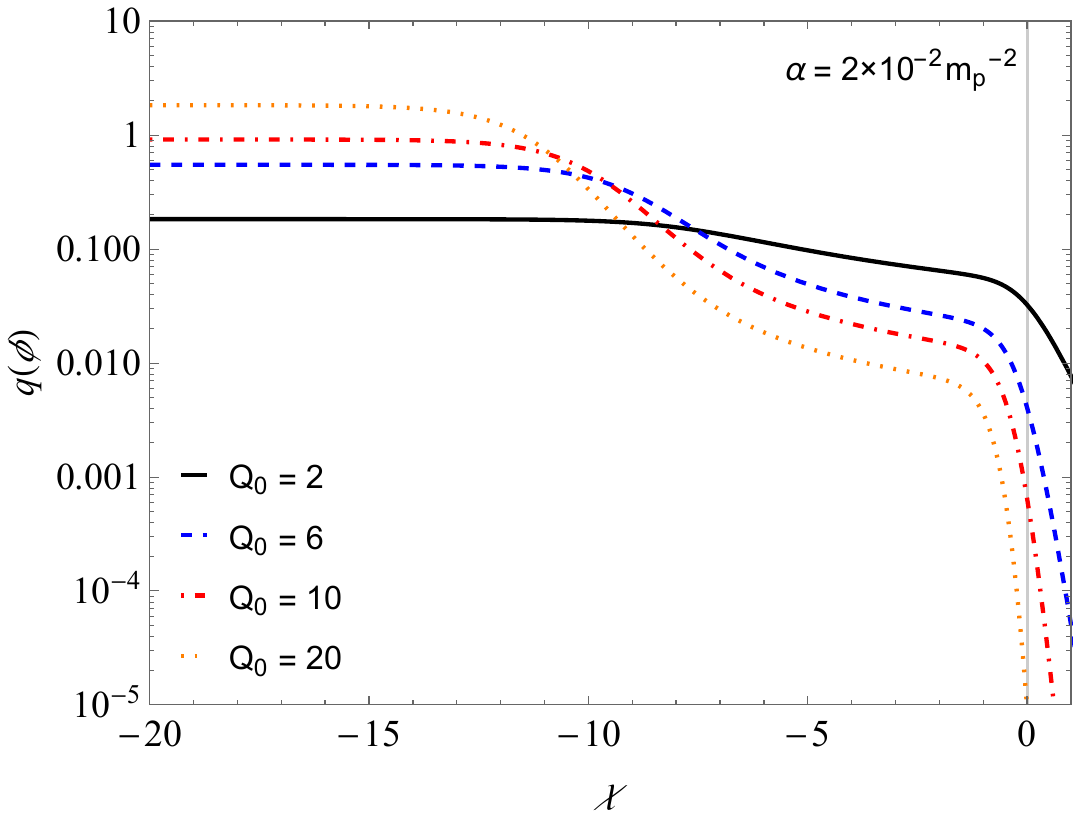}
		\caption{\small{The evolution of $Q(\phi)$ as a function of $\chi$ for fixed $\alpha=2\times10^{-2}m_{p}^{-2}$ and several values of $Q_0$.}}\label{Fig4-1}
\end{figure}

The graphic in Fig.~\ref{Fig4-1} represents the behavior of $Q(\phi)$ as a function of $\chi$ by considering different values for $Q_{0}$ and fixing $\alpha=2\times10^{ -2}m_{p}^{-2}$.
Notice that for $Q_{0}=2$ we have a very slow decay in the interval $\chi\approx[-20, 0]$, however, more pronounced outside this range. Concerning the other curves, we have a decay that becomes more accentuated with the increase of $Q_{0}$ admitting $\chi\approx[-20, -1]$. Outside this range, we have an even more significant decay.

\begin{figure}[h]
	\includegraphics[scale=0.4]{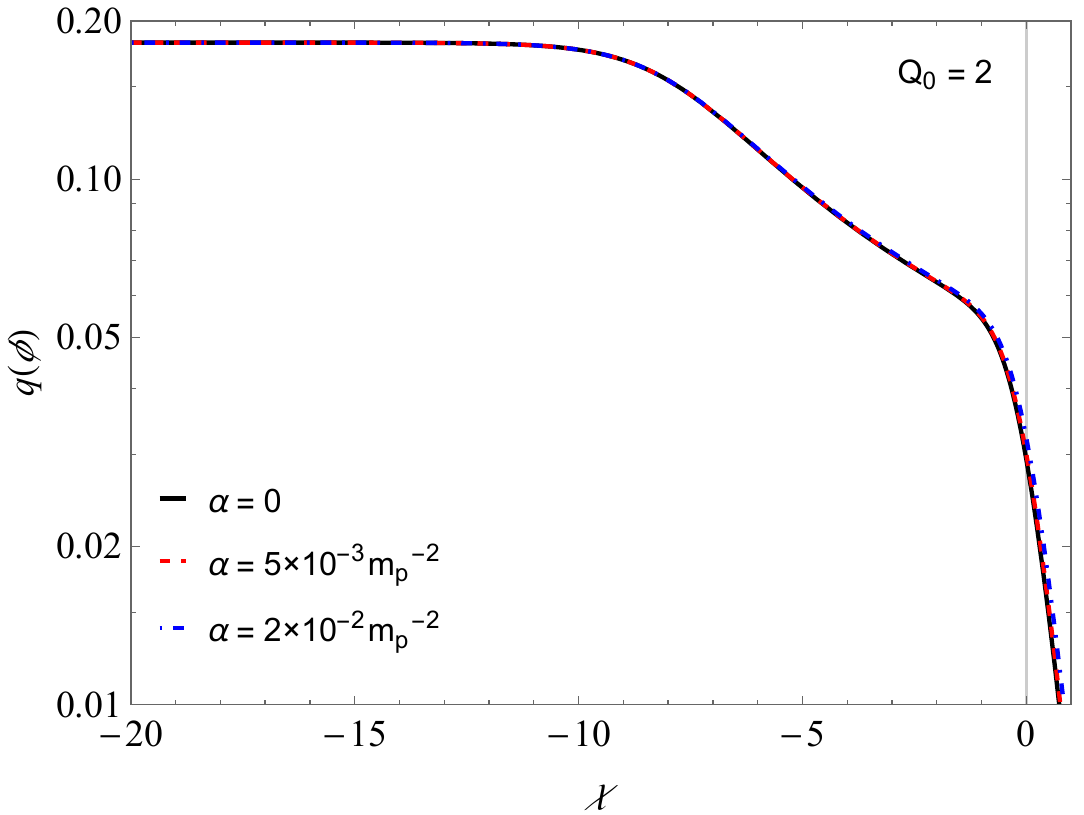}
	\includegraphics[scale=0.4]{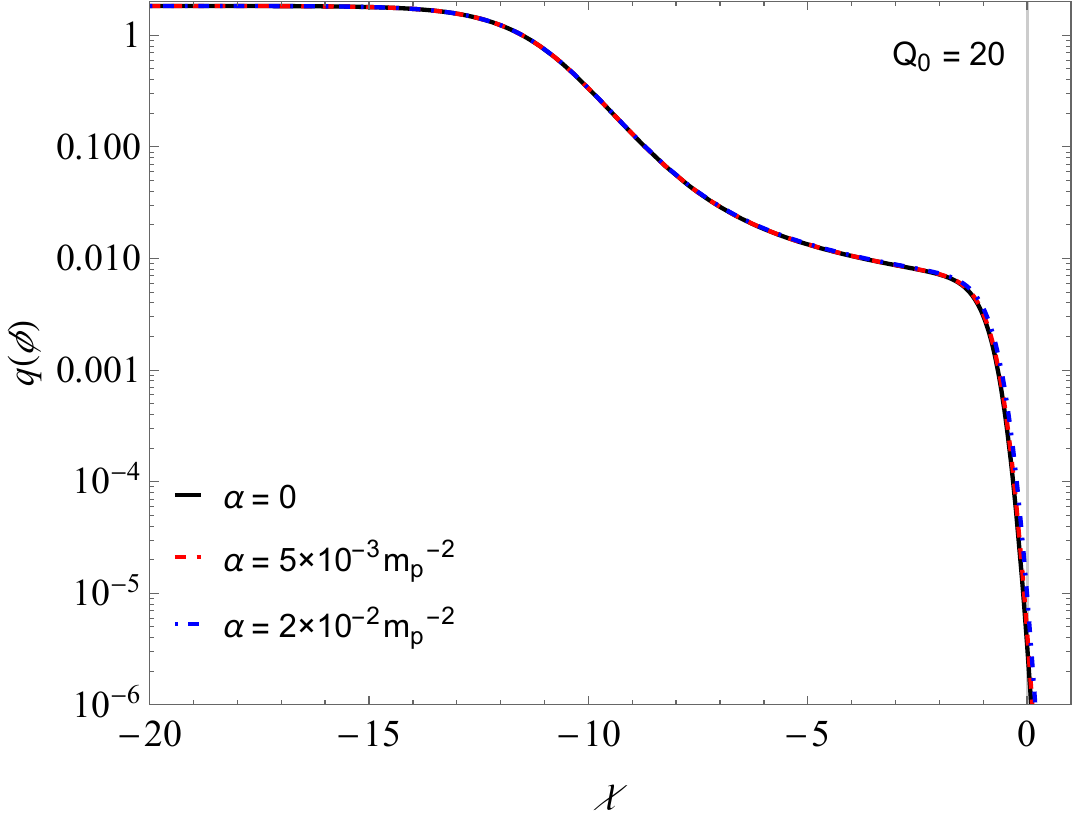}
	\caption{\small{({\it left panel}) The evolution of $Q(\phi)$ as a function of $\chi$ for fixed $Q_0=2$ and ({\it right panel}) $Q_0=20$ and several values of $\alpha$.}}\label{Fig4-2}
\end{figure}

The graphs in Fig.~\ref{Fig4-2} develops similar behavior. In the first case ({\it left panel}) we keep $Q_{0}=2$ fixed and explore some values of $\alpha$. We note that $Q(\phi)$ decays degenerately in the interval $\chi\approx[-20, -1] $, that is, without any influence of $\alpha$. Outside this range, $Q(\phi)$ decays more slowly with increasing $\alpha$. In the second case ({\it right panel}) we keep $Q_{0}=20$ fixed and also explore some values of $\alpha$ and we have a degenerate decay that becomes slower with the increase of $Q_{0}$ and without the influence of $\alpha$, in the range $\chi\approx[ -20, -1] $, however, outside this range, we have a sharp decay that increases smoothly with the decrease in $\alpha$.

\subsection{The energy densities $\rho_r$, $\rho_d$, $\rho_b$ and $\rho_\phi$}

\begin{figure}[h]
	\includegraphics[scale=0.4]{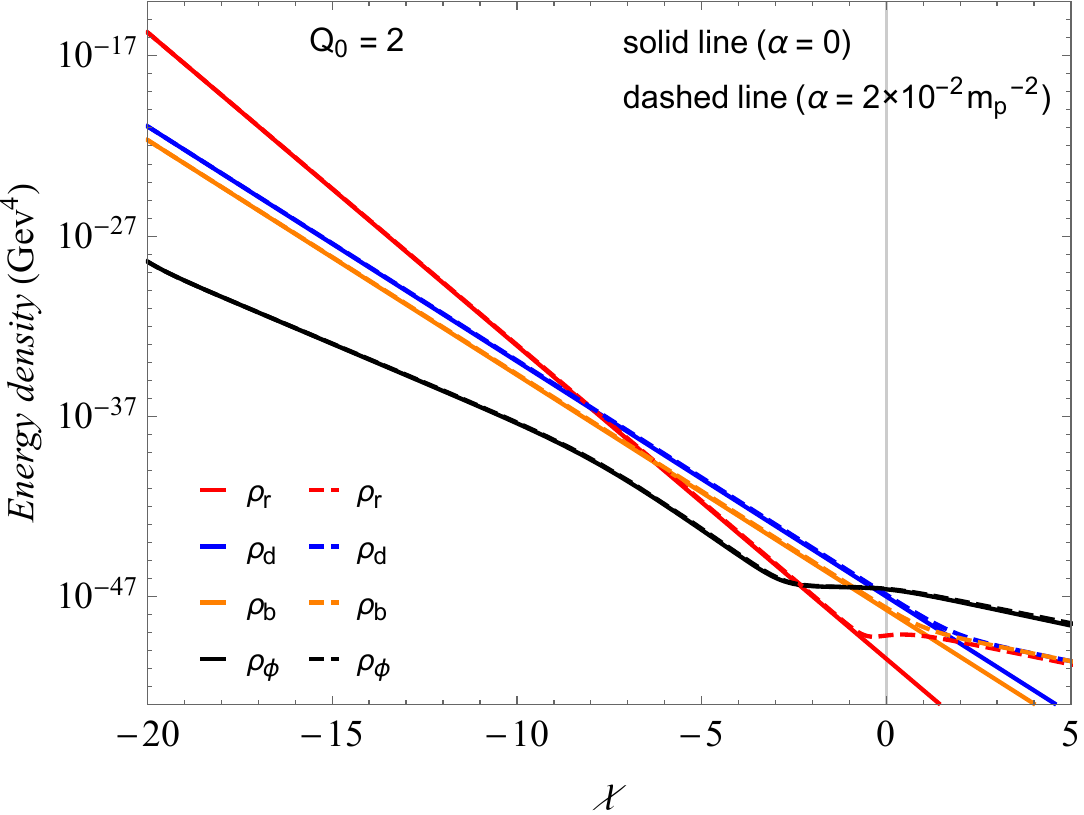}
	\includegraphics[scale=0.4]{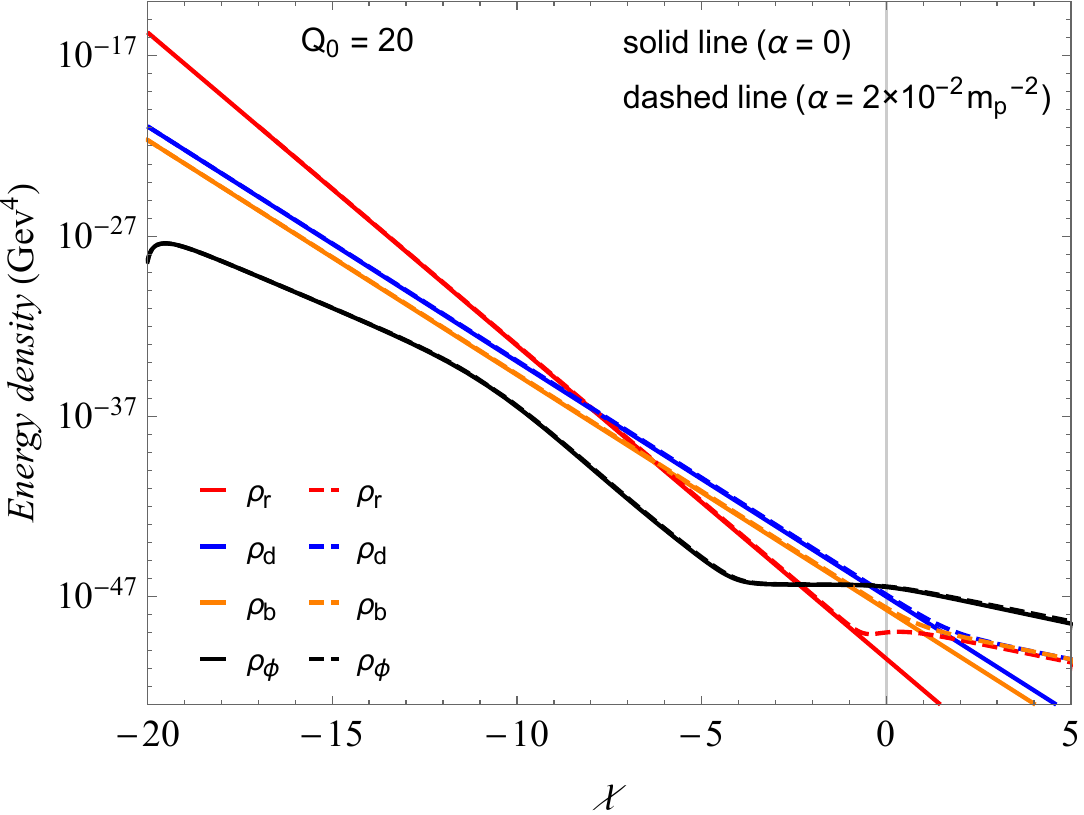}
	\caption{\small{({\it left panel}) The evolution of the energy densities for different components as a function of $\chi$ for fixed $Q_0=2$ and ({\it right panel}) $Q_0=20$ for two values of $\alpha$.}}\label{Fig5-1}
\end{figure}

The graphs in Fig.~\ref{Fig5-1} show the evolution of energy density as a function of $\chi$ for each component. In the first case ({\it left panel}) we fix $Q_{0}=2$ and consider $\alpha=0 $ and $ \alpha=2\times10^{-2}m_{p}^{-2}$. Notice that for $\alpha=0$, the components of radiation $\rho_{r}$ and dark matter $\rho_{d}$ and baryonic $ \rho_{b} $ present a linear decay, so that for $\rho_{r}$ we have a more pronounced behavior in relation to $\rho_{d}$ and $\rho_{b}$. On the other hand, assuming $\alpha=2\times10^{-2}m_{p}^{-2}$, we will have degenerate models at $\alpha=0$ in the intervals $\chi\approx[-16, -1]$, $\chi\approx[-20, 1]$ and $\chi\approx[-20, 0.5] $, for $\rho_{r}$, $\rho_{d}$ and $\rho_{b}$, respectively. Outside these ranges, each component decays more slowly, so that $\rho_{r}$, $\rho_{d}$ and $\rho_{b}$ converge their energies after $\chi\approx3$. For the dilaton component, $\rho_{\phi}$, we have a degenerate decay and non-linear scenarios, considering both $\alpha$. Such decay becomes slower after $\chi\approx -3 $, thus presenting points of intersection with each of the other components, that is, at $\chi\approx-2.5$, $ \chi\approx-1 $ and $\chi\approx-0.5$ we have the points of intersection between the decays of $\rho_{\phi}$ with $\rho_{r }$, $ \rho_{b} $ and $\rho_{d}$, respectively. Similar behavior can be found for the second case ({\it right panel}) for $Q_0=20$. 

Furthermore, in relation to the dilaton component $ \rho_{\phi} $ ({\it right panel}), it presents a more pronounced decay when compared to the graph in ({\it left panel}), that is, $ \rho_{\phi} $ has a more significant decay with the increase $ Q_{0 } $. For the points of intersection between $ \rho_{\phi} $ and the other components, we will have the same values corresponding to $ \chi $, however, such points occur at a lower energy density when compared to the graph in ({\it left panel}).

\section{Discussions}\label{discussions}

We shall first address the issue of $H_0$ tension. This problem now well known as the ``Hubble tension'' is related to the divergence of the measurements of the Hubble constant $H_0$ with respect to different applied techniques. In other words, the measurements of regions in the recent Universe such as observations from the Hubble Space Telescope of Cepheid variables 
--- see Riess et al \cite{Riess:2019cxk,Riess:2021jrx} --- present considerable different results for $H_0$ as compared with its measurements made in the early Universe by the Planck spacecraft --- for 2018 Planck release see \cite{Planck:2018vyg}. The mainly difference between these two techniques is that in the latter case the data from Planck CMB observations are processed under the base-$\Lambda$CDM model. This has raised some questions about this model and then some extended models have been put forward in the literature in order to solve the ``Hubble tension'', but according to analysis done in \cite{Planck:2018vyg} none of the extended models solves
this tension in a satisfactory way. In our present study, we also offer an alternative model to address this problem. We have considered three region of parameters to accomplish both techniques. We denominate these sets of parameters as {scenarios I, II and III} --- See Table \ref{Tb1} and Figs.~\ref{Fig1-1} and \ref{Fig2-2}.
\begin{table}[!ht]
\begin{center}
\begin{tabular}{|c|c|c|c|c|c|c|} \hline
$\rm Experiments $&$H_0$&$\alpha$&$Q_0$\\ \hline
${\rm Riess}\, et\, al.\, 2019 $&$74.03\pm1.42$ km s$^{-1}$ Mpc$^{-1} $&$-$&${\color{black}-}$\\ \hline
$\rm Planck\, 2018$&$67.3\pm1.20 $ km s$^{-1}$ Mpc$^{-1} $&$-$&$-$\\ \hline
$\rm {Scenarios} $&$H_0$&$\alpha$&$Q_0$\\ \hline
${\rm \, I}$&$67-72$ km s$^{-1}$ Mpc$^{-1} $&$0-2\times10^{-2}m_p^{-2}$&$2$\\ \hline
${\rm \, II}$&$67-74$ km s$^{-1}$ Mpc$^{-1}$&$0$&$2-20$\\ \hline
${\rm \, III}$&$68-76$ km s$^{-1}$ Mpc$^{-1}$&$5\times10^{-3} m^{-2}_p$&$2-20$\\ \hline
\end{tabular}
\end{center}
\caption{Comparison of the ranges of $ H_{0}$ values obtained by Riess et al. 2019 and Planck 2018 with Scenarios I, II, III obtained by numerical methods through the parameters $ \alpha=0 $, $ \alpha=2\times10^{-2}m_{p}^{-2} $, $ \alpha=5\times10^{-2}m_{p}^{-2} $, $ Q_{0}=2 $ and $ Q_{0}=20 $. The entire range of values of $H_{0}$ in Riess et al. 2019 is included in the range of values set out in Scenario III and partially included in the range of values in Scenario II. As for the range of $ H_{0} $ values for Planck, these partially comprise Scenarios I, II and III.
}
\label{Tb1}
\end{table}

{The Scenario I and II following the range of parameters properly chosen can simulate the results from Riess et al \cite{Riess:2019cxk,Riess:2021jrx} and Planck base-$\Lambda$CDM model. The Scenario III simulates Riess et al \cite{Riess:2019cxk,Riess:2021jrx}. Although this seems to be a reasonable test of viability of our model, a statistical analysis to find the best fit of $\alpha$ and $Q_0$ according to Planck data should be addressed elsewhere.} A similar behavior in easing the Hubble constant tension has recently appeared in the context of $W_3$ algebras \cite{Ambjorn:2021chc}.

Concerning the densities $\Omega_r$, $\Omega_d$, $\Omega_b$ and $\Omega_\phi$, depicted in Figs.~\ref{Fig1} and \ref{Fig2}, and Table \ref{Tb2}, notice that for the regimes of parameters considered in the {Scenario I, II and III}, there are no significant changes in relation to the Planck base-$\Lambda$CDM model, although yield substantial changes for the Hubble constant $H_0$. 
 
\begin{table}[]
\begin{tabular}{|c|c|c|c|c|c|c|} \hline
$\Omega_r$           & $\Omega_d$ & $\Omega_b$ & $\Omega_\phi$ & $\alpha$          & $Q_0$ \\  \hline
$9.4\times 10^{-5}$  & 0.271      & 0.049      & 0.679         & 0                 & 2     \\  \hline
$7.7\times 10^{-5}$  & 0.219      & 0.040      & 0.740         & 0                 & 20    \\  \hline
$16.2\times 10^{-5}$ & 0.302      & 0.056      & 0.636         & $2\times 10^{-2} m_p^{-2}$ & 2     \\ \hline
$19.7\times 10^{-5}$ & 0.266      & 0.047      & 0.701         & $2\times 10^{-2} m_p^{-2}$ & 20  \\ \hline                           
\end{tabular}
\caption{The table shows the values of density parameters for radiation, dark matter, baryonic matter and dilaton field (dark energy) at $\chi=0$ for different regions in the space of parameters $(\alpha, Q_0)$.}
\label{Tb2}
\end{table}
    
The dilaton field is running to achieve the limit $\phi\to-\infty$ in all scenarios discussed above, as depicted in Figs.~\ref{Fig3-1} and \ref{Fig3-2}, which guarantees the weak field field regime as we have previously assumed. Thus, the dilatonic potential \eqref{dil.potential} approaches zero at this limit as expected in quintessence scenarios. 

The initial values of the energy densities adopted above leads to acceptable current energy densities. For instance, for fixed parameters $\alpha=0, Q_0=2$, at $\chi=0$, we find 
$\rho_r=3.44\times 10^{-51}$GeV$^4$, $\rho_d=9.87\times 10^{-48}$GeV$^4$, $\rho_b=1.79\times 10^{-48}$GeV$^4$ and $\rho_{\phi}=2.47\times 10^{-47}$GeV$^4$. The matter-radiation equality occurs in the redshift $z_{eq}$. See in Table \ref{Tb3} different values for such redshift, the Hubble constant $H_0$ and $a_{eq}$ for some region of parameters.

\begin{table}[]
\begin{tabular}{|c|c|c|c|c|c|c|} \hline
$z_{eq}$ & $H_0$ & $a_{eq}$             & $\alpha$          & $Q_0$ \\ \hline
3391     & 67.0  & $2.9\times 10^{-4}$  & 0                 & 2     \\ \hline
3346.6   & 74.4  & $3.0\times 10^{-4}$  & 0                 & 20    \\ \hline
604.1    & 68.1  & $16.5\times 10^{-4}$ & $5\times 10^{-3}$ & 2     \\ \hline
425.6    & 75.5  & $23.4\times 10^{-4}$ & $5\times 10^{-3}$ & 20   \\ \hline
\end{tabular}
\caption{The table shows the acceptable values of redshift of matter-radiation equality $z_{eq}\sim 3000$ and corresponding $a_{eq}$, and the Hubble constant $H_0$ for different regions in the space of parameters $(\alpha, Q_0)$. For $\alpha\neq0$ one finds unacceptable values $z_{eq}\ll3000$. { This may be a sign that a statistical analysis can reveal that the best fit for $\alpha$ is very small or identically zero.}}
\label{Tb3}
\end{table}

\section{Conclusions}\label{conclu}

We explore a model of string theory inspired dilaton gravity in realm of modified $f(R,T)$ gravity. The numerical analyses were made in the model and revealed several cosmological quantities to describe dark energy at late time Universe. The model displays linear contributions in $T$ (trace of the energy-momentum tensor) and  { in this preliminary approach it seems to cover the well-accepted behavior of $\Lambda$CDM model for low redshifts, which is in accord with the Planck 2018 data. At this perspective,  the model also mimics extensions of the $\Lambda$CDM model due to suitable adjusted space of parameters that allows to deal with the Hubble constant $H_0$ tension.}   { We have shown in three scenarios considered in the present study that by varying appropriately some parameters such as $\alpha$ and $Q_0$ one can obtain values of $H_0$ in a wide range spanning from the Planck results to the SHOES results and beyond. The analysis on the redshift of matter-radiation equality shows better results for $\alpha=0$. This may be a sign that a statistical analysis can reveal that the best fit for $\alpha$ is very small or identically zero. Although this seems to be a reasonable test of viability of our model, a statistical analysis to find the best fit of $\alpha$ and $Q_0$ according to Planck data should be addressed elsewhere. As a perspective, we shall address these and other issues in the future.} Further studies with different models should be addressed, as for instance, one could apply several investigations on the inflationary regime such as constraining the parameters through contour plot in the plane made out of scalar spectral indices and tensor-to-scalar ratio \cite{Santos:2023zob}.

\section*{Acknowledgments}

We would like to thank CNPq, CAPES and FAPESQ-PB. J.A.V. Campos would like to thank FAPESQ-PB/CNPq n$^0$ 77/2022 for financial support and F.A. Brito acknowledges CNPq and CNPq/PRONEX/FAPESQ-PB (Grant nos. 309092/2022-1 and 165/2018). C.H.A.B. Borges also acknowledges PIQIFPB for financial support.



\begin{thebibliography}{99}


\bibitem{riess-pearlmutter} 
A. G. Riess et al., Astron. J. \textbf{116}, 1009 (1998).
S. Perlmutter et al., Astrophys. J. \textbf{517}, 565 (1999).

\bibitem{lss}
T. Koivisto, D.F. Mota, Phys. Rev. D 73, 083502 (2006).
S.F. Daniel, Phys. Rev. D 77, 103513 (2008).

\bibitem{wmap}
C.L. Bennett et al., Astrophys. J. Suppl. 148, 119-134 (2003).
D.N. Spergel et al., [WMAP Collaboration], Astrophys. J.
Suppl. 148, 175 (2003).
G. Hinshaw et al., Astrophys. J. Suppl. 208, 19 (2013).

\bibitem{cbm}
R.R. Caldwell, M. Doran, Phys. Rev. D 69, 103517 (2004).
Z.Y. Huang et al., JCAP 0605, 013 (2006).

\bibitem{bao}
D.J. Eisenstein et al., Astrophys. J. 633, 560 (2005).
W.J. Percival at el., Mon. Not. R. Astron. Soc. 401, 2148(2010).

\bibitem{ligo}
B. P. Abbott et al. (LIGO Scientific Collaboration
and Virgo Collaboration), Observation of Gravitational
Waves from a Binary Black Hole Merger, Phys. Rev. Lett. 116, 061102 (2016)

\bibitem{pad.1}
T. Padmanabhan, Dark energy and gravity, Gen. Rel. Grav. 40 (2008) 529–564.

\bibitem{frie}
J. Frieman, M. Turner, and D. Huterer, Dark Energy and the Accelerating Universe,
Ann. Rev. Astron. Astrophys. 46 (2008) 385–432.

\bibitem{martin}
J. Martin, Quintessence: a mini-review, Mod. Phys. Lett. A 23 (2008) 1252–1265.

\bibitem{cald}
R. R. Caldwell and M. Kamionkowski, “The Physics of Cosmic Acceleration,”
Ann. Rev. Nucl. Part. Sci. 59 (2009) 397–429.

\bibitem{silve}
A. Silvestri and M. Trodden, “Approaches to Understanding Cosmic Acceleration,” Rept. Prog. Phys. 72 (2009) 096901.

\bibitem{bamba}
K. Bamba, S. Capozziello, S. Nojiri, and S. D. Odintsov, “Dark energy cosmology: the equivalent description via different theoretical models and cosmography tests,” Astrophys. Space Sci. 342 (2012) 155–228.

\bibitem{li}
M. Li, X.-D. Li, S. Wang, and Y. Wang, “Dark Energy: A Brief Review,” Front. Phys. (Beijing) 8 (2013) 828–846.

\bibitem{sami}
M. Sami and R. Myrzakulov, “Late time cosmic acceleration: ABCD of dark energy and modified theories of gravity,” Int. J. Mod. Phys. D 25 no. 12, (2016) 1630031].


\bibitem{peeb}
P. J. E. Peebles and B. Ratra, “The Cosmological Constant and Dark Energy,” Rev. Mod. Phys. 75 (2003) 559–606.

\bibitem{pad.2}
T. Padmanabhan, “Cosmological constant: The Weight of the vacuum,” Phys. Rept. 380 (2003) 235–320.

\bibitem{sahni}
V. Sahni, “The Cosmological constant problem and quintessence,” Class. Quant. Grav. 19 (2002) 3435–3448.

\bibitem{velt}
H. E. S. Velten, R. F. Vom Marttens, W Zimdahl, The Eur.Phys. J. C. 74, 3160, (2014)


\bibitem{shinji}
Shinji Tsujikawa. Quintessence: A Review. Class. Quant. Grav., 30:214003,
2013.

\bibitem{buch}
Buchdahl, H. A. (1970). Non-linear Lagrangians and cosmological theory. Monthly Notices of the Royal Astronomical Society. 150: 1–8.

\bibitem{revsotirioufaraoni}
T.P. Sotiriou and V. Faraoni, $f(r)$ theories of gravity, Rev. Mod. Phys. 82 (2010) 451.
S.~Nojiri and S.~D.~Odintsov,
Unified cosmic history in modified gravity: from F(R) theory to Lorentz non-invariant models,
Phys. Rept. \textbf{505} (2011), 59-144
doi:10.1016/j.physrep.2011.04.001
[arXiv:1011.0544 [gr-qc]].
S.~Nojiri, S.~D.~Odintsov and V.~K.~Oikonomou,
Modified Gravity Theories on a Nutshell: Inflation, Bounce and Late-time Evolution,
Phys. Rept. \textbf{692} (2017), 1-104
doi:10.1016/j.physrep.2017.06.001
[arXiv:1705.11098 [gr-qc]].

\bibitem{starobin}
Starobinsky, A. A. (1980). A new type of isotropic cosmological models without singularity. Physics Letters B. 91 (1): 99–102.

\bibitem{carroll}
Carroll, S M, Duvvuri, V, Trodden, M. Turner, M S. Is cosmic speed-up due to new gravitational physics ?, Phys. Rev. D70 (2004), 043528. 
S.~Nojiri and S.~D.~Odintsov,
Modified gravity with negative and positive powers of the curvature: Unification of the inflation and of the cosmic acceleration,
Phys. Rev. D \textbf{68} (2003), 123512
doi:10.1103/PhysRevD.68.123512
[arXiv:hep-th/0307288 [hep-th]].

\bibitem{capozzi}
Capozziello, S, Cardone, V F, Carloni, S. Troisi, A “Curvature quintessence matched with observational data”, Int. J. Mod. Phys. D12 (2003), 1969-1982.


\bibitem{quantum effect}
V. Dzhunushaliev at al., Modified gravity from the quantum part of the metric, Eur. Phys. J. C \textbf{74}, 2743 (2014).
R. Yang, Effects of quantum fluctuations of metric on the universe, Physics of the Dark Universe \textbf{13}, 87 (2016).
X. Liu, T. Harko and SD. Liang, Cosmological implications of modified gravity induced by quantum metric fluctuations, Eur. Phys. J. C \textbf{76}, 420 (2016).

\bibitem{harko}T. Harko et al., Phys. Rev. D 84, 024020 (2011).


\bibitem{flrw} H. Shabani and M. Farhoudi, $f(R,T)$ cosmological models in phase space, Phys. Rev.
D \textbf{88}, 044048 (2013).

\bibitem{Min}
Min-Xing Xu, T. Harko, and Shi-Dong Liang, Quantum Cosmology of $f(R,T)$ gravity, Eur. Phys. J. C \textbf{76}, 449 (2016).

\bibitem{Moraes}
P. H. R. S. Moraes and P. K. Sahoo, The simplest non-minimal matter-geometry coupling in the $f(R,T)$ cosmology, Eur. Phys. J. C \textbf{77}, 480 (2017).

\bibitem{Shabani}
H. Shabani and A. H. Ziaie, Bouncing cosmological solutions from $f(R,T)$ gravity, Eur. Phys. J. C \textbf{78}, 397 (2018).

\bibitem{Debnath}
P. S. Debnath, Bulk viscous cosmological model in $f(R,T)$ theory of gravity, Int. J.Geom. Meth. Mod. Phys. \textbf{16}, 1950005 (2019).

\bibitem{Bhattacharjee.1}
S. Bhattacharjee and P. Sahoo, Comprehensive analysis of a non-singular bounce in $f(R,T)$ gravitation, Physics of the Dark Universe \textbf{28}, 100537 (2020).

\bibitem{Bhattacharjee.2}
S. Bhattacharjee et al., Inflation in $f(R,T)$ gravity, Eur. Phys. J. Plus \textbf{135}, 576 (2020).

\bibitem{Gamonal}
M. Gamonal, Slow-roll inflation in $f(R,T)$ gravity and a modified Starobinsky-like inflationary model, Physics of the Dark Universe \textbf{31}, 100768 (2021).


\bibitem{witten-dilaton.1}
E. Witten, Phys. Lett. B149 (1984) 351.

\bibitem{witten-dilaton.2}
E. Witten, Nucl. Phys. B443 (1995) 85.



\bibitem{dil.mod.1} M. Gasperini, Phys. Rev. D64 (2001) 043510; M. Gasperini, F. Piazza and G. Veneziano, Phys. Rev. D65 (2001) 023508.

\bibitem{Gasperini:1999ne}
M.~Gasperini,
On the response of gravitational antennas to dilatonic waves,
Phys. Lett. B \textbf{470}, 67-72 (1999)
doi:10.1016/S0370-2693(99)01309-X
[arXiv:gr-qc/9910019 [gr-qc]].

\bibitem{Damour}
T. Damour and A.M. Polyakov, The String dilaton and a least coupling principle, Nucl. Phys. B 423 (1994) 532.

\bibitem{veneziano.lageN} G. Veneziano. Large N bounds on, and compositeness limit of, gauge and gravitational interactions. JHEP 06 (2002) 051.
 
 \bibitem{Taylor}T.R. Taylor and G. Veneziano, Phys. Lett. B 213, 459 (1988).


\bibitem{riess} A. G. Riess et al. [Supernova Search Team], Observational evidence from supernovae
for an accelerating universe and a cosmological constant, Astron. J. {\bf 116}, 1009 (1998)
doi:10.1086/300499 [astro-ph/9805201].

\bibitem{Riess:2016jrr}
A.~G.~Riess, L.~M.~Macri, S.~L.~Hoffmann, D.~Scolnic, S.~Casertano, A.~V.~Filippenko, B.~E.~Tucker, M.~J.~Reid, D.~O.~Jones and J.~M.~Silverman, \textit{et al.}
A 2.4\% Determination of the Local Value of the Hubble Constant,
Astrophys. J. \textbf{826}, no.1, 56 (2016)
doi:10.3847/0004-637X/826/1/56
[arXiv:1604.01424 [astro-ph.CO]].

\bibitem{Riess:2019cxk}
A.~G.~Riess, S.~Casertano, W.~Yuan, L.~M.~Macri and D.~Scolnic,
Large Magellanic Cloud Cepheid Standards Provide a 1\% Foundation for the Determination of the Hubble Constant and Stronger Evidence for Physics beyond $\Lambda$CDM,
Astrophys. J. \textbf{876}, no.1, 85 (2019)
doi:10.3847/1538-4357/ab1422
[arXiv:1903.07603 [astro-ph.CO]].

\bibitem{Riess:2021jrx}
A.~G.~Riess, W.~Yuan, L.~M.~Macri, D.~Scolnic, D.~Brout, S.~Casertano, D.~O.~Jones, Y.~Murakami, L.~Breuval and T.~G.~Brink, \textit{et al.}
A Comprehensive Measurement of the Local Value of the Hubble Constant with 1 km s$^{-1}$ Mpc$^{-1}$ Uncertainty from the Hubble Space Telescope and the SH0ES Team,
Astrophys. J. Lett. \textbf{934}, no.1, L7 (2022)
doi:10.3847/2041-8213/ac5c5b
[arXiv:2112.04510 [astro-ph.CO]].

\bibitem{Ambjorn:2021chc}
J.~Ambjorn and Y.~Watabiki,
Easing the Hubble constant tension,
Mod. Phys. Lett. A \textbf{37}, no.07, 2250041 (2022)
doi:10.1142/S0217732322500419
[arXiv:2111.05087 [gr-qc]].

\bibitem{Planck:2018vyg}
N.~Aghanim \textit{et al.} [Planck],
Planck 2018 results. VI. Cosmological parameters,
Astron. Astrophys. \textbf{641}, A6 (2020)
[erratum: Astron. Astrophys. \textbf{652}, C4 (2021)]
doi:10.1051/0004-6361/201833910
[arXiv:1807.06209 [astro-ph.CO]].

\bibitem{Santos:2023zob}
J.~R.~L.~Santos, S.~S.~da Costa and R.~S.~Santos,
Cosmological models for f(R,T)\ensuremath{-}\ensuremath{\Lambda}(\ensuremath{\phi}) gravity,
Phys. Dark Univ. \textbf{42}, 101356 (2023)
doi:10.1016/j.dark.2023.101356


\bibitem{Poplawski}N. J. Poplawski, arXiv:gr-qc/0608031.
\bibitem{Fisch} E. Fischbach and C. Talmadge, Nature (London) 356, 207 (1992).


\end{thebibliography}
\end{document}